\documentstyle[12pt]{article}
\begin{document}
\begin{titlepage}
\hspace*{\fill} ULB--TH--94/06\\
\hspace*{\fill} NIKHEF--H 94--13\\
\hspace*{\fill} hepth 9405109

\vspace{2.5cm}
\begin{centering}

{\huge Local BRST cohomology in the antifield formalism:
I. General theorems}

\vspace{2cm}
{\large Glenn Barnich$^{1,*}$, Friedemann Brandt $^{2,**}$ and \\
Marc Henneaux$^{1,***}$}\\
\vspace{1cm}
$^1$Facult\'e des Sciences, Universit\'e Libre de Bruxelles,\\
Campus Plaine C.P. 231, B-1050 Bruxelles, Belgium\\
$^2$NIKHEF-H, Postbus 41882, 1009 DB Amsterdam,\\ The
Netherlands\\

\end{centering}

\vspace{2.5cm}

{\footnotesize \hspace{-0.6cm}($^*$)Aspirant au Fonds National de la
Recherche
Scientifique (Belgium).\\
($^{**}$)Supported by Deutsche Forschungsgemeinschaft.\\
($^{***}$)Also at Centro de Estudios
Cient\'\i ficos de Santiago, Chile.}
\end{titlepage}

\begin{abstract}
We establish general theorems on the cohomology
$H^*(s|d)$ of the BRST differential modulo the spacetime exterior
derivative,
acting in the algebra of local $p$-forms depending on the fields and
the antifields (=sources for the BRST variations).
It is shown that $H^{-k}(s|d)$ is isomorphic to $H_k(\delta |d)$ in
negative ghost
degree $-k\ (k>0)$, where $\delta$ is the Koszul-Tate differential
associated
with the stationary surface.
The cohomological group $H_1(\delta |d)$ in form degree $n$ is
proved to be isomorphic to the
space of constants of the motion, thereby providing a cohomological
reformulation
of Noether theorem. More generally, the group $H_k(\delta|d)$ in
form degree $n$
is isomorphic to the space of $n-k$ forms that are closed when the
equations of motion hold.
The groups $H_k(\delta|d)$ $(k>2)$ are shown to vanish for standard
irreducible gauge theories.
The group $H_2(\delta|d)$ is then calculated explicitly for
electromagnetism, Yang-Mills models and Einstein gravity.
The invariance of the groups $H^{k}(s|d)$
under the introduction of non minimal variables and of auxiliary
fields is also demonstrated.
In a companion paper, the general formalism is applied to the
calculation of $H^k(s|d)$
in Yang-Mills theory, which is carried out in detail for an arbitrary
compact gauge group.
\end{abstract}
\pagebreak
\def\qed{\hbox{${\vcenter{\vbox{
   \hrule height 0.4pt\hbox{\vrule width 0.4pt height 6pt
   \kern5pt\vrule width 0.4pt}\hrule height 0.4pt}}}$}}
\newtheorem{theorem}{Theorem}[sectionc]
\newtheorem{lemma}{Lemma}
\newtheorem{definition}{Definition}
\newtheorem{corollary}{Corollary}
\newcommand{\proof}[1]{{\bf Proof.} #1~$\qed$.}
\renewcommand{\theequation}{\thesection.\arabic{equation}}
\renewcommand{\thetheorem}{\thesection.\arabic{theorem}}
\renewcommand{\thelemma}{\thesection.\arabic{lemma}}
\section{Introduction}
\setcounter{equation}{0}
A major development of field theory in the eighties has been the
construction
of the antifield-antibracket formalism \cite{Batalin}. This formalism
finds its roots in earlier work on the renormalization of Yang-Mills
models \cite{Becchi,Tyutin,Zinn-Justin} and quantization of
supergravity \cite{Kallosh,deWit}, and enables one to formulate
the quantum rules (path integral, Feynman diagrams) for an
arbitrary
gauge theory in a manner that maintains manifest spacetime
covariance.

The algebraic structure of the antifield formalism has been
elucidated
in \cite{Fisch,Henneaux}, where it has been shown that the BRST
complex
contains two crucial ingredients:

(i) the Koszul-Tate resolution, generated by the antifields, which
implements
the equations of motion in (co)homology~; and

(ii) the longitudinal exterior complex, which implements gauge
invariance.

\noindent The BRST differential combines the Koszul-Tate
differential with the longitudinal
exterior derivative along the systematic lines of homological
perturbation
theory \cite{Fisch2}. As a result of that analysis, a simple rationale
for the BRST
construction has been obtained and, in particular, the role of the
antifields
has been understood. A pedagogical exposition of these ideas may be
found
in \cite{Henneaux2}.

As we have just mentioned, a key feature of the BRST differential is
that it
incorporates the equations of motion through the Koszul-Tate
resolution.
This is true both classically \cite{Fisch}, where the relevant equations
are
the classical Euler-Lagrange equations, and quantum-mechanically
\cite{Henneaux},
where the relevant equations are now the Schwinger-Dyson
equations.
[A different (non cohomological) relation between the antifields and
the
Schwinger-Dyson equations has been analyzed recently in
\cite{Alfaro}
for theories with a closed gauge algebra].

It is somewhat unfortunate that this important conceptual property
of the BRST
differential $s$ is often underplayed in the Yang-Mills context, where
what
one usually calls the BRST differential is only a piece of it, namely
the
off-shell extended longitudinal exterior derivative along the gauge
orbits.
Such a differential exists because the gauge algebra closes off-shell.
This accident of the Yang-Mills theory
(closure off-shell) hides the fundamental fact that it is the {\it full}
BRST differential $s$, including the Koszul-Tate piece $\delta$, that is of
direct
physical interest. Indeed, it is the only differential available for a
generic gauge theory. Moreover, it is the cohomology of $s$ that
appears in
renormalization theory \cite{Becchi,Zinn-Justin1} (where the
antifields are named
sources for the BRST variations), in the study of anomalies
\cite{Howe} as well
as in the question of consistently deforming the classical action
\cite{Henneaux8}.

Actually, what really appears in those problems is not just the
cohomology of $s$
but rather the cohomology of $s$ {\it in the space of local
functionals}.
The purpose of this paper is to investigate some general properties of
the cohomological
groups $H^k(s)$ of $s$ acting in the space of local functionals
with ghost number $k$, or rather,
of the related and more tractable cohomological groups $H^k(s|d)$ of
$s$
acting in the space of local $p$-forms. Here, $d$ is the exterior
derivative in spacetime.
According to homological perturbation theory, these goups are
isomorphic to
$H_{-k}(\delta |d)$ for negative $k$'s, where the subscript denotes the
antighost number, and to
$H^k(\gamma|d,H_0(\delta))$ for
positive $k$'s, where  the differential $\gamma$ is the exterior
derivative along the gauge orbits
(see below). Our main results can be summarized as follows:

(i) The group $H_{1}(\delta|d)$ in form degree $n$ is
isomorphic to the space of
non trivial conserved currents. This is actually a cohomological
reformulation of
Noether theorem. More generally, the groups $H_k(\delta|d)$ in form
degree $n$
are isomorphic to the space of non trivial $n-k$ forms
that are closed modulo the
equations of motion (``characteristic cohomology").

(ii) The groups $H_q(\delta|d)$ vanish for $q>p$ for
field theories of Cauchy
order $p$. (The ``Cauchy order" of a theory is defined below.
Usual irreducible gauge theories are of Cauchy order 2).

(iii) The complete calculation of $H_2(\delta|d)$ is carried out for
electromagnetism, Yang-Mills models and Einstein gravity. In the latter
two cases, $H_2(\delta|d)$ vanishes.  For Einstein gravity, the
vanishing of $H_2(\delta | d)$ is a consequence of the absence
of Killing vectors for a generic Einstein metric.

(iv) Non-minimal sectors, as well as the ``ultralocal" shift symmetries
of \cite{Alfaro}, do not contribute to $H(s|d)$.

(v) The invariance of $H(s|d)$ under the introduction of auxiliary
fields is established.

\noindent These general results are applied in a companion paper
\cite{Henneaux10} to the computation of  $H^k(s|d)$ for
Yang-Mills theory.

The next five sections (2 through 6) are mostly recalls of the BRST
features needed for the
subsequent analysis: how to handle locality
\cite{Voronov,Henneaux3}, examples, BRST construction
and main theorem of homological perturbation theory.
Section 7 is then devoted to the isomorphism between
$H_1(\delta|d)$ and the space of
constants of the motion. In section 8, we introduce the concept of Cauchy
order and establish some theorems
on $H_p(\delta|d)$ for theories of Cauchy order $q$. The general analysis
is pursued further in sections 9 and 10. In section 11, we prove some
general results on $H_2(\delta|d)$ for irreducible gauge theories.
These are then used in sections 12 and 13 to
compute $H_2(\delta|d)$ for Yang-Mills models.
We show that there is no $n-2$-form that is closed modulo the
equations of motion for semi-simple
gauge groups. This result holds also for Einstein gravity.
Sections 14 and 15 respectively show that non minimal sectors or
auxiliary fields do not modify
the local BRST cohomology.

We assume throughout our analysis that the topology of spacetime is
simply that of the $n$-th dimensional euclidean space ${\bf R}^n$.

\section{Cohomological groups ${\bf H^{k}(s)}$ and ${\bf H^k(s|d)}$}
\setcounter{equation}{0}
The way to incorporate locality in the BRST formalism is quite
standard and proceeds as follows.
First one observes that local functions, i.e., (smooth) functions of the
field components
and a finite number of their derivatives, are functions defined over
{\it finite dimensional
spaces}. These spaces, familiar from the theory of partial differential
equations, are called
``jet spaces" and are denoted here by $V^k\ (k=0,1,2,\dots)$.
Local coordinates on $V^k$
are given by $x^\mu$ (the spacetime coordinates), the field
components $\phi^i$, their derivatives
$\partial_\mu\phi^i$ and their subsequent derivatives
$\partial_{\mu_1\dots\mu_j}
\phi^i$ up to order $k$ ($j=0,1,2,\dots,k$). Since we assume that
spacetime is
${\bf R}^n$, these local coordinates on $V^k$ are also global
coordinates.
The Lagrangian involves usually
only the fields and their first derivatives and so is a function on
$V^1$.
We refer the reader unfamiliar with this approach to
\cite{Anderson,Olver,Saunders,McCloud} for more information.

Local functionals are by definition integrals of local functions. More
precisely, consider the
exterior algebra of differential forms on ${\bf R}^n$ with coefficients
that are local functions.
These will be called ``local $q$-forms". Local functionals are integrals
of local $n$-forms.
The second idea for dealing with locality is to reexpress all the
equations involving local
functionals in terms of their integrands. To achieve this goal, one
needs to know how to remove
the integral sign. This can be done by means of the following
elementary results:

(i) Let $\alpha$ be an exact local $n$-form, $\alpha=d\beta$. Assume
$\oint \beta =0$, where
the surface integral is evaluated over the boundary of the spacetime
region under consideration.
Then $\int \alpha = 0$ (Stokes theorem).

(ii) Conversely, if $\alpha$ is a local $n$-form such that $\int \alpha
=0$  for all allowed field
configurations, then $\alpha = d\beta$ with $\oint \beta =0$.

These results are well known and proved for instance in
\cite{Henneaux2} chapter $12$.
The differential $d$ is the exterior derivative in spacetime, defined
in the algebra of local
$q$-forms through
\begin{eqnarray}
d f(x^\mu,\phi^i, \partial_\mu \phi^i,
\dots\partial_{\mu_1\dots\mu_s}\phi^i)
=({{\partial}^R f\over\partial x^\nu}+
{{\partial}^R f\over\partial \phi^i}\partial_\nu\phi^i
\nonumber\\+{{\partial}^R f\over\partial(\partial_\mu\phi^i)}
\partial_{\nu}\partial_{\mu}\phi^i  +\dots
+{{\partial}^R f\over\partial(\partial_{\mu_1\dots\mu_s}\phi^i)}
\partial_{\nu}\partial_{\mu_1\dots\mu_s}\phi^i)dx^\nu
\end{eqnarray}
and $d(dx^\mu)=0$. As in \cite{Henneaux2}, we shall take all differentials
to act from the right.

On account of $(i)$ and $(ii)$, the correspondence between local $n$-
forms and local functionals
is not unique. If one allows only for $(n-1)$-forms such that $\oint
\beta = 0$, then, local
functionals may be viewed as equivalence classes of local $n$-forms
(which are necessarily $d$-closed)
modulo $d$-exact ones. They are thus the elements of the
cohomological space $H^n(d)$ \cite{Gel'fand}.

The BRST differential $s$ is defined in the algebra of local functions.
It may easily be extended
to the algebra of local $q$-forms by setting $s(dx^\mu)=0$. One has
\begin{equation}
sd+ds=0
\end{equation}
since $s\partial_\mu =\partial_\mu s$.
Let $A=\int \alpha$ be a BRST-closed local functional. {}From
\begin{equation}
sA=\int s\alpha=0
\end{equation}
one gets
\begin{eqnarray}
s\alpha + d\beta =0
\end{eqnarray}
with $\oint \beta =0$. Thus, the integrand $\alpha$ is a local $n$-
form that is BRST-closed modulo $d$.
Furthermore, $A$ is BRST-exact iff $\alpha=d\lambda +s\mu$ (with
$\oint \lambda =0$).
Accordingly, the cohomological groups $H^k(s)$ of $s$ acting in the
space of local functionals is isomorphic to
the cohomological groups $H^{k,n}(s|d)$ of $s$ acting in the space of
local $n$-forms
($k=$ ghost number, $n=$ form degree).

The condition that $\oint \beta$ should vanish is rather
awkward to take into account,
if only because it depends on the precise conditions imposed on the
fields at the boundaries.
For this reason, it is customary to drop it and to investigate
$H^{k,n}(s|d)$ without restrictions
on the $(n-1)$-forms at the boundary.
This approach will be followed here. By doing
so, one allows elements of
$H^{k,n}(s|d)$ that {\it do not define} $s$-closed local functionals
because of non-vanishing
surface terms. We shall comment further on this point below (end of
section 7).

\section{Regularity conditions. Examples}
\setcounter{equation}{0}
Since one can reformulate questions involving local functionals in
terms of local $n$-forms,
we shall exclusively work from now on with the algebra of local $q$-
forms. That is, any
element $a,b,c,\alpha,\beta,\dots$ upon which the differential $s$
acts will be a local
$q$-form with no restrictions at the boundaries unless otherwise
specified.

Let ${\cal L}_0={\cal L}_0(\phi^i, \partial_\mu
\phi^i,\dots,\partial_{\mu_1\dots\mu_s}\phi^i)$
be the original gauge invariant Lagrangian. The equations of motion
are
\begin{eqnarray}
{\cal L}_i \equiv {\delta {\cal L}_0\over\delta\phi^i}=0
\end{eqnarray}
where ${\delta {\cal L}_0\over\delta\phi^i}$ are the Euler-Lagrange
derivatives of ${\cal L}_0$
with respect to $\phi^i$,
\begin{eqnarray}
{\cal L}_i\equiv {\delta {\cal L}_0\over\delta\phi^i}\equiv
{\partial{\cal L}_0\over\partial\phi^i}
-\partial_\mu{\partial{\cal
L}_0\over\partial(\partial_\mu\phi^i)}+\dots +(-1)^s
\partial_{\mu_1\dots\mu_s}
{\partial{\cal L}_0\over\partial(\partial_{\mu_1\dots\mu_s}\phi^i)}.
\end{eqnarray}
Two Lagrangians ${\cal L}_0$ and ${\cal L^\prime}_0$ are regarded
as being equivalent if they
yield identical equations of motion, i.e., if they have the same Euler-
Lagrange derivatives,
${\cal L}_i\equiv{\cal L^\prime}_i$. The corresponding $n$-forms
$\alpha_0= {\cal L}_0
dx^0\dots dx^{n-1}$ and  $\alpha_0^\prime = {\cal L^\prime}_0 dx^0\dots
dx^{n-1}$ are then also called
equivalent. One has $\alpha_0 = \alpha_0^\prime +d\beta$ for some
local $(n-1)$-form $\beta$.
Conversely, if  $\alpha_0 = \alpha_0^\prime +d\beta$, then ${\cal
L}_i\equiv{\cal L^\prime}_i$ (see theorem \ref{d_cohomology}).

We shall make the same regularity assumptions on ${\cal L}_0$ and
on the gauge transformations
as in \cite{Henneaux3}. These state that one can (locally in the jet
bundle spaces) separate
the field equations ${\cal L}_i=0$ and their derivatives
$\partial_{\mu_1\dots\mu_k}
{\cal L}_i=0\ (k=1,2,\dots)$ into two groups. The first group contains
the ``independent" equations
$L_a=0$. The second group contains the dependent equations
$L_\Delta=0$,
which hold as consequences of the others. Furthermore, one may
introduce new local coordinates
in $V^k$ (for each $k$) in such a way that the independent $L_a$ are
some of the
new coordinates in the vicinity of the surface defined by the
equations of motion.
So, one can also split the field components and their derivatives into
two groups. The first group
contains the independent field variables, denoted by $x_A$, which
are not constrained by the
equations of motion in $V^k$.
The second group contains the dependent field variables, denoted by
$z_a$, which can be expressed in terms of
the $x_A$ and the $L_b$, in such a way that $z_a = z_a (x_A,L_b)$ is
smooth and invertible for the
$L_b$'s. In the case of reducible gauge theories, similar conditions
are imposed on the reducibility
functions.

These conditions are easily seen to hold for the usual gauge theories.
This is explicitely verified
in \cite{Henneaux4} for the Klein-Gordon field $\phi$ and the Yang-
Mills $A^a_\rho$.
We list here the corresponding $L_a, L_\Delta, x_A$ and $z_a$.

\noindent Klein-Gordon:
\begin{eqnarray}
\{L_a\}\equiv\{ {\cal L}\equiv\qed\ \phi,\partial_\mu {\cal
L},\partial_{\mu_1\mu_2}{\cal L},\dots\}\\
\{L_\Delta\}\ is\ empty\\
\{x_A\}\equiv\{\phi,\partial_\rho\phi,\partial_{s_1}\partial_{\rho}
\phi,\dots,\partial_{s_1\dots s_m}
\partial_\rho\phi,\dots\}\\
\{z_a\}\equiv\{\partial_{00}\phi,\partial_{\rho_1}\partial_{00}\phi,
\dots,\partial_{\rho_1\dots\rho_m 00}\phi,\dots\}\\
\end{eqnarray}
Yang-Mills:
\begin{eqnarray}
\{L_a\}\equiv\{ {\cal L}^\mu_a\equiv D_\nu
F^{\mu\nu}_a,\partial_\rho {\cal L}_a^m,
\partial_{\rho_1\rho_2}{\cal L}_a^m,
\dots,\partial_{\rho_1\dots\rho_s}{\cal L}_a^m,\dots,\nonumber\\
\partial_{k}{\cal L}_a^0,\partial_{k_1 k_2}{\cal
L}_a^0,\dots,\partial_{k_1\dots k_2}{\cal L}_a^0,\dots\}\\
\{L_\Delta\}\equiv\{\partial_{0}{\cal
L}_a^0,\partial_{\rho}\partial_0{\cal L}_a^0,
\partial_{\rho_1\rho_2}\partial_0{\cal L}_a^0,
\dots,\partial_{\rho_1\dots \rho_s}\partial_0{\cal L}_a^0,\dots\}\\
\{x_A\}\equiv\{A^a_\mu,\partial_\rho
A^a_\mu,\partial_{s_1}\partial_{\rho}A^a_m,
\dots,\partial_{s_1\dots s_k}\partial_\rho
A^a_m,\dots,\partial_\lambda\partial_0 A^a_0,\dots,\nonumber\\
\partial_{\lambda_1\dots\lambda_k}\partial_0 A^a_0,\dots,
\partial_{\bar l}\partial_m A^a_0,\dots,
\partial_{\bar l_1\dots \bar l_k}\partial_m A^a_0,\dots\},\ \ (\bar
l,\bar l_i \neq 1)\\
\{z_a\}\equiv\{\partial_{00}A^a_m,\partial_{\rho_1}\partial_{00}A^a
_m,\dots,
\partial_{\rho_1\dots\rho_s}\partial_{00}A^a_m,
\dots,\partial_{11} A^a_0,\dots,\nonumber\\
\partial_{s_1\dots s_n}\partial_{11} A^a_0\dots\}
\end{eqnarray}

In the Klein-Gordon case, for which $\{L_\Delta\}$ is empty, the set
$I_0$ of independent
variables $x_A$ enjoys a useful property: it is stable under spatial
differentiation, i.e.,
$\partial_k x_A \in I_0$ for all $k$'s and $A$'s ($\partial_k I_0
\subset I_0$).
The equations of motion constrain only the temporal derivatives of
the $x_A$'s.
This is not true for the Yang-Mills model, since $\partial_{11} A^a_0$
does not belong
to $I_0$ even though $\partial_1 A^a_0$ does. However, it is true
that $I_0$ is preserved under differentiation
with respect to $x^{\bar l}$, $\partial_{\bar l} I_0\subset I_0\ (\bar
l=2,3,\dots,n-1)$.
Furthermore, the set of independent equations $L_a$ is stable under
spatial differentiation.
We shall study the implications of these properties when we introduce
the concept of Cauchy order below (section 8).

The regularity conditions also hold for $p$-form gauge theories,
which are reducible. Namely, the
reducibility identities on the equations of motion (``Noether
identities")
\begin{eqnarray}
R_\alpha\equiv R^i_\alpha {\cal
L}_i+R^{i\mu}_\alpha\partial_{\mu}{\cal L}_i+\dots+
R^{i\mu_1\dots\mu_k}_\alpha\partial_{\mu_1\dots\mu_k}{\cal
L}_i\equiv 0\label{reducibility},
\end{eqnarray}
together with their derivatives
$\partial_{\mu}R_\alpha,\partial_{\mu_1\mu_2}R_\alpha,\dots$
fall into two  groups: the independent identities $R_u$ ; and the
dependent identities $R_U$,
which hold as consequences of $R_u=0$. [More precisely, when one
says that $R_U=0$ holds as a
consequence of $R_u=0$, one views the ${\cal L}_i$ in
(\ref{reducibility})
as independent variables not related to the $\phi^i$ ; the statement
would otherwise be meaningless].
Similar properties are verified for the higher order reducibility
functions.

For a $2$-form abelian gauge theory with gauge field $B_{\mu\nu}=-
B_{\nu\mu}$ and equations of motion
${\cal L}^{\mu\nu}\equiv\partial_\rho H^{\rho\mu\nu}=0$
($H_{\rho\mu\nu}=\partial_\rho B_{\mu\nu}+
\partial_\mu B_{\nu\rho}+\partial_\nu B_{\rho\mu}$), one has
$R^\nu\equiv\partial_\mu{\cal L}^{\mu\nu}$.
The reducibility relations $R^\nu =0$ are not independent in the sense
that they are subject to algebraic identities
$\partial_\nu R^\nu=0$ holding no matter what
${\cal L}^{\mu\nu}=-{\cal L}^{\nu\mu}$ is. This is in
contrast with the Yang-Mills case
where the relations $D_\nu {\cal L}_a^\nu\equiv 0$ are independent.
Thus, while the set $\{R_U\}$ is
empty in the Yang-Mills case, one has for the $2$-form gauge field
\begin{eqnarray}
\{R_u\}\equiv\{R^\nu\equiv\partial_\mu {\cal
L}^{\mu\nu},\partial_\lambda R^n,
\partial_{\lambda_1\lambda_2} R^n,
\dots,\partial_{\lambda_1\dots\lambda_s} R^n,\dots,\nonumber\\
\partial_l R^0,\partial_{l_1 l_2} R^0,\dots,\partial_{l_1\dots l_s}
R^0,\dots\}
\end{eqnarray}
\begin{eqnarray}
\{R_U\}\equiv\{\partial_0 R^0, \partial_\rho\partial_0
R^0,\partial_{\rho_1\rho_2}\partial_0  R^0,
\dots,\partial_{\rho_1\dots\rho_s}\partial_0  R^0,\dots\}.
\end{eqnarray}
For the equations and the field variables, the split reads
\begin{eqnarray}
\{L_a\}\equiv\{{\cal L}^{\mu\nu},\partial_{\rho}{\cal L}^{mn},
\partial_{\rho_1 \rho_2}{\cal L}^{mn},
\dots,\partial_{\rho_1\dots \rho_s}{\cal
L}^{mn},\dots,\partial_{k}{\cal L}^{0 \bar m},\nonumber\\
\partial_{k_1 k_2}{\cal L}^{0 \bar m},\dots,\partial_{k_1\dots
k_s}{\cal L}^{0 \bar m},
\dots,\partial_{\bar k}{\cal L}^{01},\partial_{\bar k_1 \bar k_2}{\cal
L}^{01},\dots,
\partial_{\bar k_1\dots \bar k_s}{\cal L}^{01},\dots\}
\end{eqnarray}
\begin{eqnarray}
\{L_\Delta\}\equiv\{\partial_0{\cal L}^{0
m},\partial_{\rho}\partial_0{\cal L}^{0 m},
\partial_{\rho_1 \rho_2}\partial_0{\cal L}^{0 m},
\dots,\partial_{\rho_1\dots \rho_s}\partial_0{\cal L}^{0
m},\dots,\nonumber\\
\partial_{1}{\cal L}^{0 1},
\partial_{k}\partial_{1}{\cal L}^{0 1},\partial_{k_1
k_2}\partial_{1}{\cal L}^{0 1},\dots,
\partial_{k_1\dots k_s}\partial_{1}{\cal L}^{0 1},\dots\}
\end{eqnarray}
\begin{eqnarray}
\{x_A\}\equiv\{B_{\mu\nu},\partial_{\rho}B_{\mu\nu},\partial_{k}
\partial_{\rho}B_{mn},
\dots,\partial_{k_1\dots
k_s}\partial_{\rho}B_{mn},\dots,\partial_{\rho}\partial_0
B_{0m},\nonumber\\
\partial_{\rho_1\rho_2}\partial_0 B_{0m},\dots,
\partial_{\rho_1\dots\rho_s}\partial_0 B_{0m},\dots,
\partial_{\bar k}\partial_{n}B_{0 \bar m},
\partial_{\bar k_1 \bar k_2}\partial_{n}B_{0 \bar
m},\dots,\partial_{\bar k_1\dots \bar k_s}
\nonumber\\ \partial_{n}B_{0 \bar m}, \dots,
\partial_{k}\partial_{1}B_{01},
\partial_{k_1 k_2}\partial_{1}B_{01},
\dots,\partial_{k_1\dots k_s}\partial_{1}B_{01},
\dots,\partial_{\hspace{.065cm}{\bar{\hspace{-.065cm}\bar
l}}}\partial_{\bar m}B_{01},\nonumber\\
\partial_{\hspace{.065cm}{\bar{\hspace{-.065cm}\bar l}}_1
\hspace{.065cm}{\bar{\hspace{-.065cm}\bar l}}_2}
\partial_{\bar m}B_{01},\dots,
\partial_{\hspace{.065cm}{\bar{\hspace{-.065cm}\bar l}}_1
\dots\hspace{.065cm}{\bar{\hspace{-.065cm}\bar l}}_s}
\partial_{\bar m}B_{01},\dots\},\ \
(\hspace{.065cm}{\bar{\hspace{-.065cm}\bar
l}},\hspace{.065cm}{\bar{\hspace{-.065cm}\bar l}}_i
=3,4,\dots,n-1).
\end{eqnarray}
\begin{eqnarray}
\{z_a\}\equiv\{\partial_{0}\partial_{0}B_{mn},\partial_{\rho}
\partial_{0}\partial_{0}B_{mn},
 \dots,\partial_{\rho_1\dots \rho_s}\partial_{0}
\partial_{0}B_{mn},\dots,\partial_{1}\partial_{1}B_{0 \bar m},
\nonumber\\
\partial_{k}\partial_{1}\partial_{1}B_{0 \bar m},
\partial_{k_1 k_2}\partial_{1}\partial_{1}B_{0 \bar m},
\dots, \partial_{k_1\dots k_s}\partial_{1}\partial_{1}B_{0 \bar
m},\dots,
\partial_{2}\partial_{2}B_{01},\nonumber\\
\partial_{\bar s}\partial_{2}\partial_{2}B_{01},
\partial_{\bar s_1 \bar s_2}\partial_{2}\partial_{2}B_{01},\dots,
\partial_{\bar s_1\dots\bar
s_n}\partial_{2}\partial_{2}B_{01},\dots\}.
\end{eqnarray}
This time, it is not true that $\partial_{\bar l}I_0\subset I_0$
for $\bar l \geq 2$. But
$\partial_{\hspace{.065cm}{\bar{\hspace{-.065cm}\bar
l}}}I_0\subset I_0$ if $\hspace{.065cm}{\bar{\hspace{-.065cm}\bar
l}} \geq 3$,
while $\partial_{\bar l}L_a\in\{L_a\}$ and $\partial_l
R_u\in\{R_u\}$.
We leave it to the reader to investigate the local structure of the
equations of motion for higher order $p$-form
gauge fields or gravity along similar lines.
Note that in all cases treated here, the $x_A$ and the independent
equations $L_a$ provide global coordinates in the jet
spaces and not just local coordinates in the vicinity of the stationary
surface. This property is clear for linear theories, but it also holds
for Yang-Mills models or gravity because the terms with highest derivatives
in $L_a$ are the linear ones. Thus, modulo terms belonging to the
previous space $V^k$, the relationship between the field derivatives of
order $k+1$ and the variables $x_A$ and $L_a$ is the same as in the
linear case. Furthermore, in the Yang-Mills case,
the change of parametrization between the field
variables and their derivatives on the one hand, and the $(x_A,L_a)$
on the other hand, is polynomial. Its inverse also fulfills this property.
Polynomiality in the original field variables and their derivatives is
accordingly completely equivalent to polynomiality in the
$x_A,L_a$. In the case of gravity, the same property is true for the
transformation restricted to variables carrying derivatives,
since the quantities $\sqrt{g}$ or $g^{\lambda\mu}$, which are non
polynomial in the
undifferentiated fields, occur in the field equations.

There is clearly a lot of freedom in the explicit choice of what is
meant by the ``independent variables" $x_A$, since any other choice
$x_A\rightarrow x^\prime_A=x^\prime_A(x_B,L_a)$ with $\partial x^\prime_A/
\partial x_B$ invertible is also acceptable. The subsequent results do not
depend on the precise choice that is being made.
All that matters is that the split
of the field variables and equations of motion with the above properties
can indeed be performed if desired.

A different split adapted to the
Lorentz symmetry - or to the $SO(n)$ symmetry in the Euclidean case -
could have been actually
achieved. This is because the Lorentz group is semi-simple. Hence, for
each $k$, the representation to which
the derivatives of order $k$ belong, is completely reducible. The
equations of motion restrict an
invariant subspace of that representation. A covariant split is
achieved by working with a basis
adopted to the irreducible subspaces of the representation of order
$k$ (for each $k$). Such
a covariant split is useful in maintaining manifest covariance.
However, the non covariant splits
given here, such that $\partial_{\bar l} I_0\subset I_0$ or
$\partial_{\hspace{.065cm}
{\bar{\hspace{-.065cm}\bar l}}} I_0 \subset I_0$,
are useful in establishing the vanishing theorems on $H(\delta|d)$
derived below. Covariance can be controled
differently, as we shall mention in section 5.

\section{Local $p$-forms and antifields}
\setcounter{equation}{0}
To fix the ideas, we shall assume from now on that the theory is at
most a reducible gauge theory
of order one and that the fields $\phi^i$ are bosonic. The gauge
transformations
\begin{eqnarray}
\delta_\varepsilon\phi^i=\int
R^i_\alpha(x,x^\prime)\varepsilon^\alpha (x^\prime) dx^\prime\\
R^i_\alpha(x,x^\prime)=R^i_\alpha\delta(x,x^\prime)+R^{i\mu}_\alpha
\partial_\mu\delta(x,x^\prime)+
\dots+R^{i\mu_1\dots\mu_k}_\alpha\partial_{\mu_1\dots\mu_k}\delta
(x,x^\prime)\\
\Longleftrightarrow\delta_\varepsilon\phi^i=R^i_\alpha
\varepsilon^\alpha+
R^{i\mu}_\alpha\partial_\mu\varepsilon^\alpha+
\dots+R^{i\mu_1\dots\mu_k}_\alpha\partial_{\mu_1\dots\mu_k}
\varepsilon^\alpha
\end{eqnarray}
are not independent
\begin{eqnarray}
\int Z^\alpha_\Delta(x,x^\prime)
R^i_\alpha(x^\prime,x^{\prime\prime})dx^\prime=0
\end{eqnarray}
but there are no non trivial relations among the $Z^\alpha_\Delta$.
The ghost and antifield spectrum relevant to that case is given by
\begin{eqnarray}
\phi^A\equiv (\phi^i,C^\alpha,C^\Delta)
\end{eqnarray}
\begin{eqnarray}
\phi^*_A\equiv (\phi^*_i,C^*_\alpha,C^*_\Delta)
\end{eqnarray}
with
\begin{eqnarray}
antigh\ \phi^i=antigh\ C^\alpha = antigh\ C^\Delta=0\\
antigh\ \phi^*_i=1,\ antigh\ C^*_\alpha=2,\ antigh\ C^*_\Delta=3\\
puregh\ \phi^i=0,\ puregh\ C^\alpha=1,\ puregh\ C^\Delta=2\\
puregh\ \phi^*_i=puregh\ C^*_\alpha=puregh\ C^*_\Delta=0\\
gh\ \phi^i=0,\ gh\ C^\alpha =1,\ gh\ C^\Delta=2\\
gh\ \phi^*_i=-1,\ gh\ C^*_\alpha=-2,\ gh\ C^*_\Delta=-3
\end{eqnarray}

Irreducible gauge theories have no $Z^\alpha_\Delta$,
and thus no ghosts of ghosts $C^\Delta$ and their antifields
$C^*_\Delta$.
Theories without gauge freedom have only $\phi^i$ and $\phi^*_i$
($R^i_\alpha\equiv 0,\ Z^\alpha_\Delta\equiv 0$).
The assumption that the gauge theory is at most reducible of order
one or that the fields are bosonic is by no means important.
It is only made to keep the formulas and the field spectrum simple.
The general theorems of sections
6,7 or 8 below hold for reducible gauge theories of higher order as
well.

The local $p$-forms introduced previously depended only on the
original fields and their derivatives.
{}From now on, local $p$-forms will also involve the ghosts, the
antifields and
their derivatives. Because the objects under consideration may have terms
of arbitrarily high
antighost number, we shall actually define two different
types of local $p$-forms and
local functionals. The $p$-forms of the first type are given by unrestricted
formal sums
\begin{eqnarray}
a=\sum_{k\geq 0}\stackrel{(k)}{a}\label{sum},
\end{eqnarray}
where $k$ is the antighost number and where $\stackrel{(k)}{a}$ are
$p$-forms of antighost number
$k$ involving the fields, the ghosts, the antifields and a finite
number of their derivatives.
Because $a$ is assumed to have given total ghost number, and
because $antigh\ \stackrel{(k)}{a}=k$,
the $\stackrel{(k)}{a}$ are actually polynomials in the ghosts and
their derivatives.
The sum (\ref{sum}) may not terminate, i.e., $a$ may be an infinite
formal series in the antifields.
So, while $\stackrel{(k)}{a}$ are local $p$-forms in the usual sense,
the formal sum
(\ref{sum}) may in principle involve derivatives of arbitrarily high
order since the order of the derivatives present in $\stackrel{(k)}{a}$
may increase with $k$. In the same way, local
functionals of the first type are given by unrestricted sums of
integrated terms
\begin{eqnarray}
A=\int\sum_{k\geq 0}\stackrel{(k)}{a}\label{intsum},
\end{eqnarray}
where $\stackrel{(k)}{a}$ are usual $n$-forms of antighost number
$k$.

For a generic gauge theory with open algebra, there is a priori no
control as to whether the
formal sums (\ref{sum}) or (\ref{intsum}) stop after a finite number
of steps. The
local $p$-forms and functionals (\ref{sum}) and (\ref{intsum}) are
accordingly the natural
objects to be considered. It is to those objects that the general
theorems of homological
perturbation theory apply.

For the usual theories like Yang-Mills models or gravity, however, it
is possible to control the
expansion (\ref{sum}) and (\ref{intsum}). For such theories, we shall
consider a second type
of local $p$-forms and local functionals, namely, those for which the
expansions
(\ref{sum}) or (\ref{intsum}) stop after a finite number of steps,
\begin{eqnarray}
a=\sum_{0\leq k\leq L}\stackrel{(k)}{a}\label{locsum}\\
A=\int\sum_{0\leq k\leq L}\stackrel{(k)}{a} \label{intlocsum}.
\end{eqnarray}
There is no difference between the individual terms appearing in the
expansion (\ref{sum}) or (\ref{locsum}). When projected to a definite value
of the antighost number, the local forms (\ref{sum}) or (\ref{locsum})
are identical. The difference
lies only in the fact that (\ref{sum}) (or (\ref{intsum})) may involve
terms of arbitrarily high antighost number.

In the case of Yang-Mills and gravity, we shall restrict even
more the functional spaces to which
the local functions and
functionals belong, by demanding that they be polynomials in the
derivatives of the fields (and
also in the undifferentiated fields in the Yang-Mills case).
That is, we exclude local functions like $exp(\partial_0 A_1)$. This is
quite natural from the point of
view of perturbative quantum field theory.

Thus, we require that the local $q$-forms be polynomials in all the variables
$\phi^i$, $C^\alpha$, $\phi^*_i$, $C^*_\alpha$ and their derivatives
for Yang-Mills models~; and for Einstein gravity, that they be
polynomials in $C^\alpha$, $\phi^*_i$, $C^*_\alpha$ and their derivatives,
as well as in the derivatives of the fields $\phi^i$, with
coefficients that may be
infinite series in the undifferentiated fields (to allow the inverse
metric $g^{\mu\nu}
=\eta^{\mu\nu}-h^{\mu\nu}+\dots$).

\section{BRST differential}
\setcounter{equation}{0}

The Koszul-Tate differential $\delta$ is defined in the algebra of local
$p$-forms by:
\begin{eqnarray}
\delta \phi^A(x)=0,\
\delta\phi^*_i(x)=-{\delta{\cal L}_0\over\delta\phi^i}(x)\nonumber\\
\delta  C^*_\alpha(x)=\int \phi^*_i(x^\prime) R^i_\alpha
(x,x^\prime)dx^\prime\nonumber\\
\delta C^*_\Delta(x)= \int -C^*_\alpha (x^\prime) Z^\alpha_\Delta
(x,x^\prime)dx^\prime
+{1\over 2} \int\phi^*_i (x^\prime)\phi^*_j (x^{\prime\prime}) M^{ij}_\Delta
(x;x^\prime x^{\prime\prime})dx^\prime dx^{\prime\prime}\nonumber\\
\delta dx^\mu=0
\label{eqmotion}
\end{eqnarray}
where $\phi^A$ denotes collectively $\phi^i$, $C^\alpha$ and $C^\Delta$.
[Like $s$ and $\gamma$ below, the differential $\delta$ is extended to the
derivatives of the fields by requiring $\delta\partial_\mu
=\partial_\mu\delta$, and to arbitrary functions of the generators by
means of the Leibnitz rule]. It has antighost number $-1$ and is such that
\begin{eqnarray}
H_0(\delta)=C^\infty(\Sigma)\otimes {\bf C} [C^\alpha,\partial_\mu
C^\alpha,\dots,
C^\Delta,\partial_\mu C^\Delta,\dots,dx^\mu]\nonumber\\
H_k(\delta)=0,\ k>0
\end{eqnarray}
where $C^\infty(\Sigma)$ denotes the quotient algebra of the smooth
functions of the fields $\phi^i$
and their derivatives modulo the ideal of functions that vanish when
the field equations hold.
One says that $\delta$ provides a ``resolution" of the algebra
$C^\infty(\Sigma)\otimes {\bf C}[C^\alpha,
\partial_\mu C^\alpha,\dots,C^\Delta,\partial_\mu C^\Delta,\dots,dx^\mu]$,
the antighost number being the
resolution degree. The same result holds if instead of arbitrary smooth
functions of the fields and their derivatives, one considers polynomial
functions, provided that the change of variables $(\phi^i,\partial\phi^i,
\partial^2\phi^i,\dots)\leftrightarrow (x_A,L_a)$ and its inverse are both
polynomial. Any polynomial cycle $a$ of antighost
number $k>0$ can be written as $a=\delta b$ where $b$ is also polynomial
(\cite{Henneaux6} footnote 1).

The acyclicity of $\delta$ in strictly positive antighost number is
most easily proved by introducing
a homotopy for the operator that counts the antifields and the
equations of motion \cite{Henneaux2},
for instance in the basis of section 3. Since that basis is not
manifestly covariant, one may
wonder whether acyclicity also holds in the space of Lorentz invariant
local forms. More generally,
if the theory is invariant under a global symmetry group G, one may
wonder whether acyclicity also
holds in the space of local forms belonging to a definite
representation of G. That this is so if the
group is semi-simple can be seen either by redefining the basis of
section 3 in a manner compatible
with the symmetry, or by using the fact that $\delta$ commutes with
the action of G
and hence maps any irreducible
representation occuring in the decomposition
of the completely reducible representation of G given by the local
forms on an equivalent representation
or on zero \cite{Henneaux6}. [The argument uses the fact that the
representations of G are completely
reducible~; this is guaranteed to hold if G is semi-simple, but holds
also if G is compact provided
(in both cases) that the representations under consideration are
effectively finite-dimensional. This
is for instance so if the local forms are polynomials in the fields and
their derivatives].

The differential $\gamma$ has antighost number equal to
zero. It is defined on the fields through
\begin{eqnarray}
\gamma \phi^i(x)=\int R^i_\alpha (x,x^\prime) C^\alpha (x^\prime)
dx^\prime\nonumber\\
\gamma C^\alpha(x) = \int Z^\alpha_\Delta (x,x^\prime) C^\Delta
(x^\prime)dx^\prime +{1\over 2}\int
C^\alpha_{\beta\gamma}(x;x^\prime x^{\prime\prime})
C^\beta (x^\prime)C^\gamma
(x^{\prime\prime}) dx^\prime dx^{\prime\prime}\nonumber\\
\gamma C^\Delta(x) = \int C^\Delta_{\alpha\Gamma} (x;x^\prime
x^{\prime\prime})
C^\alpha (x^\prime) C^\Gamma (x^{\prime\prime})dx^\prime dx^{\prime\prime}
\nonumber\\+{1\over 3}\int
C^\Delta_{\alpha\beta\gamma}(x;x^\prime x^{\prime\prime}
x^{\prime\prime\prime})C^\alpha
(x^\prime)C^\beta (x^{\prime\prime})C^\gamma (x^{\prime\prime\prime})
dx^\prime dx^{\prime\prime} dx^{\prime\prime\prime}
\end{eqnarray}
so that $H^*(\gamma)$ in $C^\infty(\Sigma)\otimes {\bf C}[C^\alpha,
\partial_\mu C^\alpha,\dots,C^\Delta,\partial_\mu C^\Delta,\dots]$ is
isomorphic to the cohomology
of the exterior derivative along the gauge orbits.
Furthermore, it is extended to the antifields in such a way that it is a
differential modulo $\delta$, i.e.,
\begin{eqnarray}
\gamma\delta +\delta\gamma=0\nonumber\\
\gamma^2=-(\delta s_1 +s_1\delta)
\end{eqnarray}
for some derivation $s_1$ of antighost number $+1$
\cite{Henneaux2}.

The following theorems are standard results of the BRST formalism.
\begin{theorem}\label{hpt}{\em{\bf :}}
There exists a derivation $s$ of total ghost number equal to $1$ such
that

(i) $s=\delta +\gamma+s_1+$``higher orders", antigh (``higher
orders")$\geq 2$~;

(ii) $s^2 =0$ ($s$ is a differential).
\end{theorem}
Furthermore, one may choose $s$ so that it is canonically generated
in the antibracket, $sa=(a,S)$, where
$S$ is a solution of the classical master equation $(S,S)=0$ and starts
like $S=\int {\cal L}dx +\int\dots$.
The differential $s$ is the BRST differential.
\begin{theorem}[on the cohomology of s]\label{s_cohomology}{\em{\bf :}}
The cohomology of the BRST differential $s$ in the algebra of local
$q$-forms is given by

(i) $H^k(s)=H_{-k}(\delta)=0$ for $k<0$

(ii) $H^k(s)\simeq H^k(\gamma,H_0(\delta))$ for $k\geq 0$

where $H^k(\gamma,H_0(\delta))$ is the cohomology of $\gamma$ in
the cohomology of $\delta$.
\end{theorem}
Furthermore, the correspondence between $H^k(s)$ and
$H^k(\gamma,H_0(\delta))$ is given by
\begin{eqnarray}
[a]\in H^k(s)\longleftrightarrow[a_0] \in H^k(\gamma,H_0(\delta))
\end{eqnarray}
where $a_0$ is the component of $a$ of antighost number zero.
{\it That is, for non negative ghost number, any cohomological class of the
BRST cohomology is completely determined by its antifield
independent component, which is a solution
of $\gamma a_0 +\delta a_1 =0$ or, what is the same $\gamma
a_0\approx 0$}. Here, $\approx$ means ``equal modulo the equations
of motion".

\proof{The proofs of theorem \ref{hpt} and theorem
\ref{s_cohomology} may be found
in \cite{Henneaux2}.}

We stress again that here, $s$ and $a$ are a priori infinite formal
power series
with terms of arbitrarily high antighost numbers.

To analyze the cohomology of $s$ modulo $d$, we shall also need the
following two results
\begin{theorem}[on the cohomology of d]
\label{d_cohomology}{\em{\bf :}}
The cohomology of $d$ in the algebra of local $p$-forms is given by
\begin{eqnarray}
H^0(d)\simeq {\bf R},\nonumber\\
H^k(d)=0\ for\ k\neq 0,\ k\neq n,\nonumber\\
H^n(d)\simeq  space\ of\ equivalence\ classes\ of\ local\ n-forms,
\end{eqnarray}
where two local $n$-forms $\alpha = f dx^0\dots dx^{n-1}$ and
$\alpha^\prime=f^\prime dx^0\dots dx^{n-1}$
are equivalent if and only if $f$ and $f^\prime$ have identical Euler-
Lagrange derivatives with respect to
all the fields and antifields,
\begin{eqnarray}
{\delta (f-f^\prime)\over\delta\phi^A} =0={\delta (f-
f^\prime)\over\delta\phi^*_A}\Longleftrightarrow
\alpha \ and\ \alpha^\prime\ are\ equivalent.
\end{eqnarray}
\end{theorem}
\proof{The proof of this theorem may be found in various different
places
\cite{Vinogradov,Brandt,DuboisViolette,Anderson,Wald,Dickey}.}

\begin{theorem}[on the cohomology of ${\bf \delta}$ modulo
$d$]\label{deltad_cohomology}{\em{\bf :}}
In the algebra of local forms,
\begin{eqnarray}
H_k(\delta|d)=0
\end{eqnarray}
for $k>0$ {\underbar {and}} pureghost number $>0$.
\end{theorem}

\proof{see \cite{Henneaux3} or \cite{Henneaux2}, chapter 12.}

Theorem \ref{d_cohomology} is sometimes referred to as the
``algebraic Poincar\'e lemma" because it reminds one of the usual
Poincar\'e
lemma. However, it is not the standard Poincar\'e lemma, which
states that $d\Psi=0\Rightarrow
\Psi=d\chi$ locally in spacetime but without guarantee that $\chi$
involves the fields and a
finite number of their derivatives if $\Psi$ is a local $p$-form.
Note also that $H^k(d)$ vanishes for $k\neq 0$ and $k\neq n$ only if
one allows for an explicit coordinate
dependence of the local forms. Otherwise, $H^k(d)$ is isomorphic to
the set of constant forms
for $k \neq 0$ and $k\neq n$.

\section{Homological Perturbation Theory and $H^*(s|d)$}
\setcounter{equation}{0}
\setcounter{theorem}{0}
Theorem \ref{s_cohomology} relates the cohomology of $s$ to the
cohomology of $\delta$ and $\gamma$. This is done
through the methods of homological perturbation theory. A
straightforward application of the
same techniques enables one to characterize the cohomology of $s$
modulo $d$.
\begin{theorem}[on the cohomology of $s$ modulo
$d$]\label{sd_cohomology}{\em{\bf :}}
\begin{eqnarray}
(i) H^k(s|d)\simeq H_{-k}(\delta|d)\ for\ k<0\\
(ii)H^k(s|d)\simeq H^k(\gamma|d,H_0(\delta))\ for\ k\geq 0
\end{eqnarray}
\end{theorem}

\proof{the proof proceeds as the proof of \ref{s_cohomology} (see
\cite{Henneaux2}, chapter 8, section 8.4.3).
We shall thus only sketch it here. Let $a$ be a representative of a
cohomological class of $s$ modulo $d$,
$sa+db=0$. Assume $gh\ a =k$. Expand $a$ according to the antighost
number,
\begin{eqnarray}
a=a_i+a_{i+1}+\dots\label{antifddec}\\
antigh\ a_j=j\geq 0\\
pure\ gh\ a_j = k+j\\
gh\ a_j = k
\end{eqnarray}
The first term $a_i$ in (\ref{antifddec}) has antighost number equal
to max ($0,-k$) (i.e., $i=0$
if $k\geq 0$ and $i=-k$ if $k<0$). If $k\geq 0$, $a_0$ fulfills
$\gamma a_0+\delta a_1+db_0=0$ and
thus, defines an element of $H^k(\gamma|d,H_0(\delta))$ (both $a_0$
and $b_0$ fulfill $\delta a_0
=\delta b_0 =0$ ; furthermore, $pure\ gh\ a_0=k$). It is easy to
verify that the map $H^k(s|d)
\longrightarrow H^k(\gamma|d,H_0(\delta)):[a]\longmapsto [a_0]$ is
well defined, i.e., does not depend on
the choice of representatives. One proves that it is injective and
surjective as in \cite{Henneaux2}
using the crucial property that $H_j(\delta|d)$ vanishes for both $j>0$ {\it
and} strictly positive pure
ghost number (the pure ghost number of the higher order terms in the
expansion (\ref{antifddec}) is
$>0$).
Turn now to the case $k=-k^\prime<0$. Then the expansion of $a$
reads
\begin{eqnarray}
a=a_{k^\prime}+a_{k^\prime+1}+\dots \label{s2}
\end{eqnarray}
The term $a_{k^\prime}$ fulfills $\delta a_{k^\prime}+d b_{k^\prime-
1}=0$, i.e., defines an element of
$H_{k^\prime}(\delta|d)$. As for $k\geq 0$, the map  $H^k(s|d)\longrightarrow
H_{-k}(\delta|d):
[a]\longmapsto [a_{-k}]$ is well defined and is both injective and
surjective thanks
to the triviality of $H_j(\delta|d)$ in positive antighost and pure
ghost numbers. This proves the
theorem. Note again that the series (\ref{antifddec}) or (\ref{s2})
under consideration may be infinite
formal series in the antifields, just as in theorem
\ref{s_cohomology}~; there is at this stage no
guarantee that they stop.}

Comments:

(i) For $k\geq 0$, the cohomological classes of $H^k(s|d)$ are
completely determined by their antifield independent
components. In particular, to determine whether there  exist non
trivial elements of $H^k(s|d)$,
it is enough to determine whether there exist non trivial solutions  of
$\gamma a_0+db_0+\delta a_1=0$,
or, what is the same, $\gamma a_0+db_0\approx 0$. This is true for
any value of the (positive)
ghost number, in particular for $k=1$ (anomalies). It is just the
transcription, in terms of local
functionals, of standard and well-established properties of $H^k(s)$.
Theorem \ref{sd_cohomology}
is discussed along the same lines in exercise 12.9 of
\cite{Henneaux2}.

(ii) For $k<0$, the cohomological classes of $H^k(s|d)$ are also
determined by their components of
lowest antighost number. In this case, these components do involve
the antifields but do not involve the ghosts.

(iii) The surjectivity of the map $[a]\mapsto [a_{-k}]$ for $k<0$
shows that any solution $a_{k^\prime}$
of $\delta a_{k^\prime}+db_{k^\prime-1}=0$ is automatically
annihilated by $\gamma$ up to
$\delta$- and $d$- exact terms ($\gamma a_{k^\prime}=-\delta
a_{k^\prime +1}-db_{k^\prime}$). That is,
any solution of $\delta a_{k^\prime}+db_{k^\prime-1}=0$ is ``weakly
gauge invariant" up to
$d$-exact terms.

\section{Constants of the motion and ${\bf H^n_1(\delta|d)}$.}
\setcounter{equation}{0}
\setcounter{theorem}{0}
 Although $H_j(\delta)$ vanishes for antighost number $j>0$, this is
not true for $H_j(\delta|d)$. A counterexample was provided
in \cite{Henneaux3}. In this section, we characterize more completely
$H^n_1(\delta\d)$ (where
$n$ is the form degree). We show that there is a bijective
correspondence between $H_1^n(\delta|d)$
($\equiv H^{-1,n}(s|d)$) and the space of non trivial conserved
currents. That the BRST cohomology
involves the constants of the motion is not surprising, in view of the
fact that the BRST differential
incorporates explicitly the equations of motion.

Elements of $H^n_1(\delta|d)$ are determined by $n$-forms of
antighost number one solving
\begin{eqnarray}
\delta a +dj=0\label{symmetryeq}
\end{eqnarray}
where $j$ is a $(n-1)$-form of antighost number zero. Both $a$ and
$j$ may be assumed not to
depend on the ghosts (ghost dependent contributions are trivial, see
theorem \ref{deltad_cohomology}). If one substitutes
\begin{eqnarray}
a=a^i\phi^*_i+a^{i\mu}\partial_\mu
\phi^*_i+\dots+a^{i\mu_1\dots\mu_t}\partial_{\mu_1\dots\mu_t}
\phi^*_i
\label{symmetry}
\end{eqnarray}
in (\ref{symmetryeq}), one gets using (\ref{eqmotion}),
\begin{eqnarray}
a^i{\delta {\cal L}_0\over\delta \phi^i}+a^{i\mu}\partial_\mu {\delta
{\cal L}_0\over\delta \phi^i}
+\dots+a^{i\mu_1\dots\mu_t}\partial_{\mu_1\dots\mu_t}{\delta
{\cal L}_0\over\delta \phi^i}
=dj ,
\end{eqnarray}
or, in dual notations ($a\equiv Xdx^0\dots dx^{n-1}$),
\begin{eqnarray}
X^i{\delta {\cal L}_0\over\delta \phi^i}+X^{i\mu}\partial_\mu {\delta
{\cal L}_0\over\delta \phi^i}
+\dots+X^{i\mu_1\dots\mu_t}\partial_{\mu_1\dots\mu_t}{\delta
{\cal L}_0\over\delta \phi^i}
=\partial_\mu j^\mu \label{dualsymmetry}.
\end{eqnarray}
Thus, $j^\mu$ is a current that is conserved by virtue of the
equations of motion.

The current $j$ is not completely determined by (\ref{symmetryeq}).
One may add to it an arbitrary solution
$\bar j$ of $d\bar j =0$ without changing $a$. Since $H^{n-1}(d)=0$
(theorem \ref{d_cohomology}), $\bar j$ is of the
form $dk$. Thus given $a$, $j$ is determined up to $j\rightarrow
j+dk$. But $a$ is not even
given completely~;
what is fixed is the cohomological class of $a$ in $H^n_1(\delta|d)$,
i.e., $a$ up to $\delta m+dn$.
The modification $a\rightarrow a+\delta m +dn$ yields $j\rightarrow
j+\delta n +dk$. Since
$j$ is $d$ closed modulo $\delta$, there is accordingly a well defined
map from $H^n_1(\delta|d)$
to $H^{n-1}_0(d|\delta)$,
\begin{eqnarray}
H^n_1(\delta|d)\longrightarrow H^{n-1}_0(d|\delta), [a]\longmapsto
[j]
\end{eqnarray}
The map is injective because $H_1(\delta)=0$ (if $[j]=0$ in
$H(d|\delta)$, i.e., $j=\delta n +dk$,
then $\delta (a -dn )=0$, i.e., $a=dn+\delta b$, i.e., $[a]=0$ in
$H(\delta|d)$).
It is also clearly surjective. Thus there is the isomorphism
\begin{eqnarray}
H^n_1(\delta|d)\simeq H^{n-1}_0(d|\delta)\label{isomorphism}.
\end{eqnarray}
This result is a particular case of a proposition (Eq.(16)) of
\cite{DuboisViolette}.

To fully appreciate the physical content of Equation
(\ref{isomorphism}), one needs to introduce the
concept of non trivial conserved currents and non trivial global
symmetries.

(i) A current is said to be identically conserved if it is conserved
independently of the dynamics, i.e.,
if $dj=0$ or $j=dk$. A conserved current is said to be non trivial if it
does not coincide on-shell
with an identically conserved current, $j\not\approx dk$. The space
$H^{n-1}_0(d|\delta)$ is
just the space of inequivalent non trivial conserved currents.

(ii) By making integrations by parts if necessary, one may assume
that $a\equiv
Xdx^0\dots dx^{n-1}$ does not involve the derivatives of the
antifields.
With $X=X^i\phi^*_i$, Eq.(\ref{dualsymmetry}) reduces to
\begin{eqnarray}
X^i{\delta {\cal L}_0\over\delta \phi^i}=\partial_\mu j^\mu
\end{eqnarray}
and shows that $a$ defines the symmetry $\delta_X\phi^i=X^i$ of the
action
(any linear function of the $\phi^*_i$ is naturally viewed as a tangent
vector
to field space \cite{Witten}).
Gauge symmetries (including on-shell trivial symmetries
\cite{Henneaux2})
are physically irrelevant since two configurations differing by a
gauge symmetry
must be identified. They correspond to $X^i$ of the form
\begin{eqnarray}
X^i(x)=\int  R^i_\alpha(x,x^\prime)t^\alpha(x^\prime)dx^\prime +
\int \mu^{ij}(x,x^\prime){\delta S_0\over\delta\phi^j(x^\prime)}
dx^\prime
\end{eqnarray}
with $\mu^{ij}(x,x^\prime)=-\mu^{ji}(x,x^\prime)$, which is equivalent to
\begin{eqnarray}
X^i\phi^*_i =\delta \mu+\partial_\mu b^\mu
\end{eqnarray}
[e.g., if the gauge transformations are $\delta_\varepsilon\phi^i
=R^i_\alpha
\varepsilon^\alpha + R^{i\mu}_\alpha\partial_\mu
\varepsilon^\alpha$,
then $\delta C^*_\alpha = R^i_\alpha\phi^*_i - \partial_\mu
(R^{i\mu}_\alpha
\phi^*_i)$. If
$X^i=R^i_\alpha\lambda^\alpha+R^{i\mu}_\alpha\partial_\mu
\lambda^\alpha
+\mu^{ij}{\delta {\cal L}_0\over\delta \phi^j}$ for some
$\lambda^\alpha(\phi,\partial\phi,\dots)$
and $\mu^{ij}(\phi,\partial\phi,\dots)=-
\mu^{ji}(\phi,\partial\phi,\dots)$,
then, $X^i\phi^*_i=\delta(\lambda^\alpha C^*_\alpha-{1\over 2}
\mu^{ij}\phi^*_i\phi^*_j)+\partial_\mu b^\mu$ with $b^\mu=
\lambda^\alpha
R^{i\mu}_\alpha\phi^*_i$]. A symmetry of the action is said to be a
nontrivial
global symmetry if it does not coincide with a gauge symmetry
modulo on-shell
trivial symmetries, i.e., if it is not of the form $\delta\mu +
\partial_\mu b^\mu$.
Thus, one can identify $H^n_1(\delta|d)$ with the space of
inequivalent global
symmetries.

We have therefore established
\begin{theorem}{\em{\bf :}}
\label{symmetries}
The space of inequivalent non trivial conserved currents is
isomorphic with
the space of non trivial inequivalent global symmetries,
\begin{eqnarray}
H^{n-1}_0(d|\delta)\simeq H^n_1(\delta|d)\label{basiso}
\end{eqnarray}
\end{theorem}

This theorem provides a cohomological reformulation of the physical
version of Noether theorem.
It shows that each non trivial symmetry defines a non trivial
conserved charge and vice-versa.

Comments:

(i) Given a conserved current, one may define a conserved charge through
$\int j^0 d^{n-1}x$. Now, the antibracket induces a map
$H^n_1(\delta|d) \times H^n_1(\delta|d)\longrightarrow H^n_1(\delta|d)$,
which corresponds to the Lie bracket of the global symmetries
(the antibracket coincides with the Schouten bracket \cite{Witten,Henneaux2},
which reduces to the Lie bracket for vector fields). It is easy to
verify that the isomorphism (\ref{basiso}) associates with the
Lie bracket of two global symmetries the Poisson bracket
of the corresponding conserved charges. This will be made more precise in
\cite{Henneaux11}.

(ii) The fact that gauge symmetries lead to ``trivial" conserved
currents is well
known, see e.g. \cite{Jackiw} in that context. Our analysis
reformulates the question
as a cohomological problem.

(iii) One may wonder how the non triviality of $H^n_1(\delta|d)$ in
the space
of local functions is compatible with the triviality of
$H_1(\delta,{\cal A})$ in the space ${\cal A}$
of all (local and non-local) functionals. It turns out that this follows
from a
combination of two features:
$(\alpha)$ non trivial elements of  $H^n_1(\delta|d)$ do not
necessarily define elements
of $H_1(\delta)$ upon integration because of non vanishing surface
terms~;
$(\beta)$ those that do, actually turn out to define trivial elements
that can be
written as the $\delta$-variation of non local functionals. [If this is
not the
case, then there are missing global antifields for antifields, since the
triviality of $H_1(\delta)$ is a central property of the BRST
formalism that
fixes the antifield spectrum].

To see this, let us assume that the spacetime volume under
consideration
is limited by two spacelike hypersurfaces ``at $t_1$ and $t_2$".
Let us also assume that the field
configurations are not restricted at $t_1$ and $t_2$, so that the
allowed
histories include all the solutions of the equations of motion and not
just one.
Then a solution $a$ of the equation $\delta a + dj =0$ defines a
solution
$A=\int a$ of $\delta A =0$ iff $\oint j = Q(t_2)-Q(t_1)=0$ with
$Q(t)=\int_{\Sigma(t)} j^0 d^{n-1} x$. This is a strong condition on $j$.
Indeed, the requirement $Q(t_2)-Q(t_1)=0$ for all allowed field
configurations
at $t_2$ and $t_1$ generically implies $Q(t)=constant$ and thus
$j^0=\partial_k S^{0k}$.
Thus, $\tilde j^\mu
=j^\mu -\partial_\rho S^{\mu\rho}$ with $S^{mn}=0$ and $S^{k0}=
-S^{0k}$, is a current such that $(i)$ $\tilde j^0 =0$ ; and $(ii)$ $\delta
a+d\tilde j=0$. The corresponding charge $Q(t)=\int \tilde j^0 d^{n-1}
x$
identically vanishes. Thus the transformations of the fields associated
with it
is a gauge symmetry (exercise 3.3 (b) of \cite{Henneaux2}). If all
gauge symmetries have
been properly taken into account, locally and non locally, then, $\int
a = \delta X$.
As an example, one may consider electromagnetism with
$a=A^{*0}dx^0\dots dx^{n-1}$.
One has $\delta A^{*0} +\partial_\mu j^\mu=0$ with $j^\mu =F^{\mu
0}$, i.e.,
$j^0=0$, $j^k=F^{k0}$. Thus $A = \int A^{*0}dx^0\dots dx^{n-1}$ solves
$\delta A =0$.
Even though $a$ is a non trivial element of $H(\delta|d)$ in the space
of local
functions, one has $A^{*0}(t,\vec x)=\int^t[-\delta C^*(u,\vec x) +
\partial_k
A^{*k}(u,\vec x)] du$ and thus $A=\delta \int dt\dots dx^{n-1}\int^t
-C^*(u,\vec x) du$:
$A$ is a trivial element of $H_1(\delta)$ if one includes non local
functionals.

(iv) There exist interesting theories for which there is no non trivial,
local
conserved current. For example, pure Einstein gravity is such a
theory \cite{Torre}.
In that case, the cohomological groups $H^n_1(\delta|d)$ and thus
also
$H^{-1,n}(s|d)$ are empty.

\section{Results on ${\bf H^p_k(\delta|d)}$ (p, k arbitrary)}
\setcounter{equation}{0}
\setcounter{theorem}{0}

The above theorem characterizes $H^n_1(\delta|d)$ in terms of
conserved currents. What is the
cohomology of $\delta$ modulo $d$ for the other values of the
antifield number and form degree?
As a first step in characterizing $H^p_k(\delta|d)$ for arbitrary $k$'s
and $p$'s we establish

\begin{theorem}[descent equations for $\delta$ and
$d$]\label{descent_eq_deltad}{\em{\bf :}}
if $p\geq 1$ and $k>1$, then
\begin{eqnarray}
H^p_k(\delta|d)\simeq H^{p-1}_{k-1}(\delta|d)\label{descent_iso}.
\end{eqnarray}
\end{theorem}
\proof{from $\delta a_k^p+da_{k-1}^{p-1}=0$, one gets $d\delta a_{k-
1}^{p-1}=0$ and thus using the triviality of
$d$ in degree $p-1$,
\begin{eqnarray}
\delta a_{k-1}^{p-1}+da_{k-2}^{p-2}=0
\end{eqnarray}
(If $p-1=0$, $H^0(d)$ is trivial because $a^0_{k-1}$ has non vanishing
antifield number and cannot be constant).
This shows that $a_{k-1}^{p-1}$ defines an element of $H^{p-1}_{k-
1}(\delta|d)$. It is easy to check
that this element does not depend on the choices of representatives
and thus, there is a well defined
map from $H^{p}_{k}(\delta|d)$ to $H^{p-1}_{k-1}(\delta|d)$. This
map is injective because
$H_{k}(\delta)=0$ and surjective because $H_{k-1}(\delta)=0$. This
proves the theorem.}

Of course one has also, by the same techniques as in the previous
section
\begin{theorem}\label{gen_sym}{\em{\bf :}}
if $p\geq 1$ and $k\geq 1$ with $(p,k)\neq (1,1)$, then
\begin{eqnarray}
H^p_k(\delta|d)\simeq H^{p-1}_{k-1}(d|\delta)\label{inversion}
\end{eqnarray}
Furthermore,
\begin{eqnarray}
H^1_1(\delta|d)\simeq H^0_0(d|\delta)/{\bf R}.
\end{eqnarray}
\end{theorem}
[If one does not allow for an explicit $x$ dependence in the local
forms, then, (\ref{inversion})
must be replaced by $H^p_1(\delta|d)\simeq H^{p-
1}_{0}(d|\delta)/\{constant\ forms\}$ for $k=1$].

In particular, $H^n_k(\delta|d)\simeq H^{n-k}_{0}(d|\delta)$: the
equivalence classes of $n$-forms
that are $\delta$-closed modulo $d$ at antighost number $k$ are in
bijective correspondence with the
equivalence classes of antifield independent $(n-k)$-forms that are
$d$-closed modulo the equations
of motion.

The calculation of the general solution of $da\approx 0$, $antigh\
a=0$, is a question that is of
interest independently of the BRST symmetry. It can be analyzed
without ever introducing the antifields
or the Koszul-Tate resolution and carries the name of ``characteristic
cohomology" \cite{Bryant1}. However, as we shall see in the explicit
case of the Yang-Mills
theory, the direct calculation of $H^n_k(\delta|d)$ may be simpler
than that of $H^{n-k}_0(d|\delta)$
for $k=2$. Thus, it appears to be useful to bring in the tools of the
antifield formalism even in the analysis
of questions that are a priori unrelated to the BRST symmetry, like
that of calculating
$H_0^p(d|\delta)$.

A direct consequence of theorem \ref{descent_eq_deltad} is that
$H^p_k(\delta|d)$ vanishes whenever
$k>p$. Indeed, by using repeatedly (\ref{descent_iso}), one gets
$H^p_k(\delta|d)
\simeq H^0_{k-p}(\delta|d)\simeq H^0_{k-p}(\delta)\simeq 0$
$(k>p)$.

To determine the cohomological groups $H^p_k(\delta|d)$, it is
enough to compute $H^n_k(\delta|d)$
for $k=1,2,\dots,n$ or $H^p_1(\delta|d)$ for $p=1,2,\dots,n$. In
general, this is a difficult task.
For theories of Cauchy order $q$, however, one can locate more
precisely the values of the degrees where the
non trivial cohomology may lie.

To define the Cauchy order of a theory, we come back to
the split of
the field components, the field equations
and their derivatives performed in section 3. We recall that the set of
independent field variables
$x_A$ was denoted by $I_0$. We shall say that the split has Cauchy order
$q$  if
$\partial_\alpha I_0 \subset I_0$ for $\alpha = q, q+1,\dots,n-1$.
This terminology is motivated by the fact that the split of the derivatives
is somewhat adapted to the Cauchy problem.
Thus, the above split for the Klein-Gordon theory has Cauchy order
$1$~; the one for electromagnetism and Yang-Mills theories, has Cauchy order
$2$~; and the one for $p$-form
gauge fields, has Cauchy order $p+1$.

As such the Cauchy order depends on the choice of
$I_0$ but also on the coordinate
system. For instance, the two-dimensional Klein-Gordon equation
$\partial^\mu\partial_\mu\phi=0$
reads in light-like coordinates $\partial_+\partial_-\phi=0$. One may
take as independent
field variables $\phi$, $\partial_-^{(k)}\phi$ and $\partial_+^{(k)}\phi$
($k=1,2,3,\dots$).
These are, however, preserved neither by $\partial_+$ nor by
$\partial_-$, so that the
value of $q$  associated with this choice is 2. We shall define the
Cauchy order of
a theory as the minimum value of $q$ for which $\partial_\alpha
I_0\subset I_0$ ($\alpha=q,q+1,\dots$).
The minimum is taken over all sets of spacetime coordinates and all
choices of $I_0$.

So, the Cauchy orders of the Klein-Gordon
theory and electromagnetism are respectively $\leq 1$ and $\leq 2$.
The fact that $H_1(\delta|d)$ (respectively $H_2(\delta|d)$) does not
vanish for those models implies, however, that $q=1$
(respectively $q=2$). Indeed, one has

\begin{theorem}\label{rest_norm_theo}{\em{\bf :}}
for theories of Cauchy order $q$,
\begin{eqnarray}
H^i_1(\delta|d)=0\ if\ i\leq n-q \label{van}
\end{eqnarray}
Thus, for the Klein-Gordon model, only $H^n_1(\delta|d)$ is non
vanishing. For electromagnetism
and Yang-Mills, only $H^n_1(\delta|d)$ and $H^{n-
1}_1(\delta|d)\simeq H^n_2(\delta|d)$ may be non vanishing.
And for $p$-form gauge fields, only $H^n_1(\delta|d)$, $H^{n-
1}_1(\delta|d)\simeq H^n_2(\delta|d)$,
\dots up to $H^{n-p}_1(\delta|d)\simeq H^{n-
p+1}_2(\delta|d)\simeq\dots\simeq H^n_{p+1}(\delta|d)$
may differ from zero.
\end{theorem}
\proof{We set $d = \bar d + \hspace{0.14cm}{\bar{\hspace{-
0.14cm}\bar d}}$ where $\bar d =\partial_0 dx^0 +\partial_1
dx^1+\dots
+\partial_{q-1} dx^{q-1}$ and $\hspace{0.14cm}{\bar{\hspace{-
0.14cm}\bar d}} = \partial_q
dx^q+\partial_{q+1}dx^{q+1}+\dots+\partial_{n-1}
dx^{n-1}$. Let $a$ be a solution of $\delta a + db =0$ with $antigh\
a=1$ and $deg\ a =i\leq n-q$.
One has $antigh\ b=0$ and $deg\ b=i-1$. Let $a=a_1 +a_2$ and
$b=b_1+b_2$, where $a_1$ (respectively
$b_1$) involves at least one $dx^\beta$ with $\beta\leq q-1$, while
$a_2$ (respectively
$b_2$) involve only the $dx^\alpha$ with $\alpha\geq q$. One may
assume without loss of generality that
$b_2$ involves only the independent variables $x_A\in I_0$. This
can always be achieved by adding to
$b$ a $\delta$-exact term if necessary.This modifies $a$ by a $d$-exact
term. The equations $\delta
a+db=0$
splits as
\begin{eqnarray}
\delta a_1+db_1+\bar d b_2=0\nonumber\\
\delta a_2 +\hspace{0.14cm}{\bar{\hspace{-0.14cm}\bar d}} b_2=0.
\end{eqnarray}
Now, $\hspace{0.14cm}{\bar{\hspace{-0.14cm}\bar d}} b_2$ contains
only the variables not constrained
by the equations of motion since $\partial_\alpha I^0\in I^0$ for
$\alpha =q,q+1,\dots,n-1$,
while $\delta a_2$ vanishes by the equations of motion. Hence
$\hspace{0.14cm}{\bar{\hspace{-0.14cm}\bar d}} b_2$ and $\delta
a_2$
must be zero separately. This implies $a_2=\delta m_2$ and $b_2
=\hspace{0.14cm}{\bar{\hspace{-0.14cm}\bar d}} c_2$ because
$H^{i-1}(\hspace{0.14cm}{\bar{\hspace{-0.14cm}\bar d}})=0$ ($b_2$
is a $(i-1)$-form in the $(n-q)$-dimensional space of the
$x^q,x^{q+1},\dots,x^{n-1}$~; and $i\leq n-q$ by assumption).
Thus, by making the redefinitions
$a\rightarrow a -\delta m$ and
$b\rightarrow b-dc$, we may assume $a=a_1$ and $b=b_1$.

To pursue the analysis, we split further $a_1$ and $b_1$ into
components $a_{11}$ ($b_{11}$)
involving at least two $dx^\beta$ with $\beta\leq q-1$ and $a_{12}$
($b_{12}$) involving only
one $dx^\beta$ ($\beta \leq q-1$). We further redefine $b_{12}$ in
such a way that it involves only
$x_A$, $b_{12}\rightarrow b_{12}+\delta t$. Because $t$ involves
one $dx^\beta$, the corresponding redefinition
of $a$ ($a\rightarrow a-dt$) leaves $a_2$ equal to zero. The equation
$\delta a +db =0$ yields
\begin{eqnarray}
\delta a_{12} +\hspace{0.14cm}{\bar{\hspace{-0.14cm}\bar d}}
b_{12}=0.
\end{eqnarray}
from which one infers as above that $\delta a_{12}=0$ and
$\hspace{0.14cm}{\bar{\hspace{-0.14cm}\bar d}}
b_{12}=0$. It is easy to see that
this implies not only $a_{12}=\delta m_{12}$ but also
$b_{12}=\hspace{0.14cm}{\bar{\hspace{-0.14cm}\bar d}} c_{12}$
(write
$b_{12}$ as $\sum^{q-1}_{\beta =0}b_{12\beta}dx^\beta$ where
$b_{12\beta}$ are $(i-2)$-forms
which must separately fulfill $\hspace{0.14cm}{\bar{\hspace{-
0.14cm}\bar d}} b_{12\beta}=0$. Thus one can remove $a_{12}$ from
$a_1$ and $b_{12}$ from $b_1$. By going on in the same fashion, one
arrives in maximal form degree
for $\bar d$ at $\delta a_{1 q}+\hspace{0.14cm}{\bar{\hspace{-
0.14cm}\bar d}} b_{1 q}=0$. Again both terms have to vanish
separately and
can be absorbed by redefinitions, which proves the theorem.
Note that (\ref{van}) holds also in the space of polynomial
$q$-forms if the change of parametrization $(\phi^i,\partial\phi^i,\dots)
\leftrightarrow (x_A,L_a)$ and its inverse are polynomial.}

An analogous vanishing theorem for the characteristic cohomology
of an exterior differential system,
which probably encompasses theorem \ref{rest_norm_theo}, has
been derived in \cite{Bryant1}.

Comment:

As a side comment, we note that the result (\ref{van}) extends to the
other rows of the variational bicomplex\footnote{We thank Niky Kamran for
asking us this question.}: by theorem \ref{gen_sym},
theorem\ref{rest_norm_theo} is equivalent to the statement
that $H^i_0(d|\delta)=0$ for $i\leq n-q-1$
for theories of Cauchy order $q$.
This corresponds, in the terminology of the variational bicomplex for
differential equations \cite{Anderson}, to the exactness of the bottom row
of this complex up to horizontal degree $n-q-1$.

Now, it is straightforward to check that the same result remains true
for any other row with vertical degree $s$ different from zero.
Indeed, in the proof of theorem \ref{rest_norm_theo}, one has to replace
$b$ by a linear combination of $b$'s multiplied by s of the generators
$d_V\phi^i,d_V\phi^i_{,\mu},d_V\phi^i_{,\mu_1\mu_2},\dots$,
whereas $a$ becomes a linear combination of $a$'s linear in the antifields
and their derivatives multiplied by $s$ of the above generators plus a
linear combination of antifield independent $a$'s multiplied by
$s-1$ of the above generators and one of the generators $d_V\phi^*_i,
d_V\phi^*_{i,\mu},d_V\phi^*_{i,\mu_1\mu_2},\dots$. Using the split of the
fields and their derivatives into $L_a$ and $x_A$, one can choose
$b$ to depend only on $x_A$ and $d_V x_A$. Indeed, both $L_a$ and its
vertical derivative $d_V L_a$ can be removed from $b$ since they are
$\delta$-exact ($d_V \delta + \delta d_V =0$). Since $d_V$ and $d\equiv d_H$
(respectively $\hspace{0.14cm}{\bar{\hspace{-0.14cm}\bar d}}$)
anticommute, $\hspace{0.14cm}{\bar{\hspace{-0.14cm}\bar d}}b$ will also
only depend on $x_A$ and $d_V x_A$. Because
$H^i(\hspace{0.14cm}{\bar{\hspace{-0.14cm}\bar d}})=0$ in vertical
degree $s$, the proof of theorem \ref{rest_norm_theo} goes through
exactly in the same way.

This means that the variational bicomplex for differential equations of
Cauchy order $q$ is exact up to order $n-q-1$ (all the columns for this
bicomplex are exact like in the case of the free complex
\cite{Anderson}). This question will be developed further in
\cite{Henneaux11}.

\section{Linear gauge theories}
\setcounter{equation}{0}
\setcounter{theorem}{0}

The vanishing theorem \ref{rest_norm_theo} for the $\delta$-cohomology
modulo $d$ can be derived, in the case of linear gauge theories and
perturbations of them (in a sense to be made precise), under different
conditions. The techniques necessary for deriving this alternative
vanishing theorem are quite useful and based on the well known fact that
the Euler-Lagrange derivatives of a divergence $\partial_\mu j^\mu$
identically vanish.

\begin{theorem}\label{9.1}{\em{\bf :}}
for a linear gauge theory of reducibility order $r$, one has,
\begin{eqnarray}
H^n_j(\delta|d)=0,\qquad j>r+2
\end{eqnarray}
whenever $j$ is strictly greater than $r+2$ (we set $r=-1$ for a theory
without gauge freedom).
\end{theorem}

\proof{assume for definiteness $r=0$ (irreducible gauge theory). The case
of arbitrary $r$ is treated along identical lines. Since the equations
are linear, one has
\begin{eqnarray}
{\delta^L {\cal L}\over\delta\phi^i}=D_{ij}\phi^j
\end{eqnarray}
where $D_{ij}$ is a linear differential operator with field independent
coefficients,
\begin{eqnarray}
D_{ij}=\sum_{l\geq 0}d_{ij}^{\mu_1\dots\mu_l}\partial_{\mu_1\dots\mu_l}.
\end{eqnarray}
Similarily, the Noether identity reads
\begin{eqnarray}
U^i_\alpha{\delta^L {\cal L}\over\delta\phi^i}=0\label{9.4}
\end{eqnarray}
with
\begin{eqnarray}
U^i_\alpha=\sum_{k\geq 0}u_\alpha^{i\mu_1\dots\mu_k}
\partial_{\mu_1\dots\mu_k}.
\end{eqnarray}

Let $a$ be a $n$-form solution of $\delta a + \partial_\mu b^\mu =0$
(in dual notations), with $antighost\ a \geq 3$. By taking the
Euler-Lagrange derivatives of this cycle condition with respect to
$C^*_\alpha$, $\phi^*_i$ and $\phi^i$, one gets ($\delta (\partial_\mu
b^\mu)/\delta(anything)=0$),
\begin{eqnarray}
\delta {\delta^R a\over\delta C^*_\alpha}=0,\qquad
\delta {\delta a^R\over\delta\phi^*_i}-
U^{i+}_\alpha{\delta^R a\over\delta C^*_\alpha}=0\nonumber\\
\delta {\delta^R a\over\delta\phi^i}-
D^{+}_{ji}{\delta^R a\over\delta\phi^*_j}=0
\end{eqnarray}
where
\begin{eqnarray}
U^{i+}_\alpha=\sum_{k\geq 0}(-)^k u_\alpha^{i\mu_1\dots\mu_k}
\partial_{\mu_1\dots\mu_k}\\
D^{+}_{ij}=\sum_{l\geq 0}(-)^l
d_{ij}^{\mu_1\dots\mu_l}\partial_{\mu_1\dots\mu_l}.
\end{eqnarray}
Since the variational derivatives of $a$ have non-vanishing antighost
number ($antigh\ a\geq 3$), the relation $H_k(\delta)=0$ ($k>0$)
implies, using the operator identity $U^i_\alpha D_{ij}=0$ that follows
from (\ref{9.4}),
\begin{eqnarray}
{\delta^R a\over\delta C^*_\alpha}=\delta f^\alpha\label{9.10}\\
{\delta^R a\over\delta\phi^*_i}=U^{i+}_\alpha f^\alpha +\delta f^i\\
{\delta^R a\over\delta\phi^i}=D^{+}_{ji}f^j+\delta f_i\label{9.12}
\end{eqnarray}
for some $f^\alpha$, $f^j$ and $f_i$. The equations
(\ref{9.10})-(\ref{9.12}) are valid for any field configuration.
Thus, we may replace in them the fields, the antifields and their
derivatives by $t$ times themselves, where $t$ is a real parameter.
For instance,
$({\delta a\over\delta C^*_\alpha})(t)=(\delta f^\alpha)(t)$
with $F(t)\equiv F(t\phi^i,t\phi^*_i,tC^*_\alpha)$.

Now, one can reconstruct $a$ from its Euler-Lagrange
derivatives through the formula
\begin{eqnarray}
a=\int_0^1[{\delta^R a\over\delta C^*_\alpha}(t)C^*_\alpha
+{\delta^R a\over\delta\phi^*_i}(t)\phi^*_i+
{\delta^R a\over\delta\phi^j}(t)\phi^j]dt +\partial_\mu k^\mu. \label{9.13}
\end{eqnarray}
If one inserts (\ref{9.10})-(\ref{9.12}) in (\ref{9.13}) one gets,
using the fact that $\delta$ does not depend on $t$ because the
equations of motion are linear
($(\delta x)(t)=\delta (x(t))$), that the cycle $a$ is given by
\begin{eqnarray}
a=\delta [(\int_0^1f^\alpha(t)dt)C^*_\alpha-
(\int_0^1f^i(t)dt)\phi^*_i+(\int_0^1f_i(t)dt)\phi^i]
+\partial_\mu k^{\prime\mu}.\label{9.14}
\end{eqnarray}
That is, $a$ is $\delta$-trivial modulo $d$ as claimed above.}

\section{Normal theories}
\setcounter{equation}{0}
\setcounter{theorem}{0}

Theorem \ref{9.1} can be extended to non linear theories under the
condition that the linear part of the theory contains the maximum number
of derivatives. We shall call such theories ``normal theories".
We shall first illustrate the concept in the case
of the Yang-Mills field coupled to coloured multiplets, and we
shall then define it in general.

The Lagrangian for the Yang-Mills field coupled to matter reads
\begin{eqnarray}
{\cal L}={1\over 8}tr F^{\mu\nu}F_{\mu\nu}+{\cal L}^y (y^i,D^y_\mu y^i)
\end{eqnarray}
with
\begin{eqnarray}
D^y_\mu y^i=\partial_\mu y^i-gA^a_\mu T^i_{aj}y^j,\\
F^{a}_{\mu\nu}=\partial_\mu A^a_\nu -\partial_\nu A^a_\mu
-g C^a_{bc} A^b_\mu  A^c_\nu.
\end{eqnarray}
Here, the $T_a$'s are the generators of the representation to which the
matter fields $y^i$ belong. We shall consider for definiteness
the case of Dirac fermions, $y^i\equiv (\psi^i,\bar\psi_i)$,
\begin{eqnarray}
{\cal L}^y=\bar\psi_i\gamma^\mu D_\mu \psi^i +m^i_j \bar\psi_i\psi^j,
\end{eqnarray}
where $m^i_j$ is the mass matrix, which commutes with $T_a$,
$[m,T_a]=0$, and we have absorbed a factor $i$ in the definition of
$\gamma^\mu$.

The Koszul-Tate differential is given by
\begin{eqnarray}
\delta A^a_\mu=0,\ \delta C^a=0,\ \delta y^i=0,\nonumber\\
\delta A^{*\mu}_a=-D_\nu F^{\nu\mu}_a + g j^\nu_a,\
\delta C^*_a=-D_\mu A^{*\mu}_a + g T^j_{ai}y^*_j y^i,\nonumber\\
\delta \bar\psi^{*i}=-\gamma^\mu D_\mu \psi^i-m^i_j\psi^j,\
\delta\psi^{*}_{i}=-D_\mu \bar\psi_i\gamma^\mu+\bar\psi_j m^j_i.
\end{eqnarray}
One can split both the Lagrangian and the Koszul-Tate differential
into free and interacting pieces,
\begin{eqnarray}
{\cal L}={\cal L}^{free}+{\cal L}^{int},\\
\delta=\delta^{free}+\delta^{int}.
\end{eqnarray}
A crucial feature of the free Lagrangian is that it contains
the maximum number of derivatives, namely, two derivatives of $A^a_\mu$ and
one derivative of $\psi^i$. The interaction vertices have at most one
derivative of $A^a_\mu$ and no derivative of $\psi^i$.

To formalize this property, we introduce a degree $K$ defined as
\begin{eqnarray}
K = N_{\partial} + A
\end{eqnarray}
where $N_{\partial}$ is the derivation counting the number of derivatives of
the fields and of the antifields,
\begin{eqnarray}
N_{\partial}=\sum_{A,(k)} {\partial^R\over\partial \phi^{A(k)}} |k|
\phi^{A(k)},
\end{eqnarray}
and where $A$ is defined by
\begin{eqnarray}
A= \sum_{(k)} {\partial^R\over\partial
(\partial^{(k)}A^{*\mu}_a)}2 \partial^{(k)}A^{*\mu}_a
+ \sum_{(k)}
{\partial^R\over\partial (\partial^{(k)}C^*_a)}3\partial^{(k)}C^*_a\nonumber\\
+\sum_{(k)} {\partial^R\over\partial
(\partial^{(k)}y^*_i)} \partial^{(k)}y^*_i.
\end{eqnarray}
The differential $\delta$ splits into components of definite $K$-degree,
\begin{eqnarray}
[K, \delta^j] = j \delta^j.
\end{eqnarray}
Since we have assigned $A$-weight $2$ to the antifields associated with
second order equations of motion and $A$-weight one to the antifields
associated with first order equations of motion, the differential
$\delta$ contains only components of non positive $K$-degree.  Explicitly,
one has
\begin{eqnarray}
\delta = \delta^0 + \delta^{-1} + \delta^{-2}
\end{eqnarray}
Similarly, one gets
\begin{eqnarray}
\delta^{free} = \delta^{free,0} + \delta^{free, -1}.
\end{eqnarray}
The derivation $\delta^{free,0}$ is simply the mass-independent
piece of $\delta^{free}$,\\
 $(\delta^{free})_{m=0} = \delta^{free,0}$.
In addition, the zeroth component of $\delta$ coincides with the zeroth
component of $\delta^{free}$ since the free part of the equations
of motion contains the maximum number of derivatives.

The differential $\delta^{free,0} \equiv \delta^0$, like
$\delta^{free}$ and $\delta$, is acyclic at positive antighost number.
Thus, if the local $q$-form $a$ (i) has positive antighost number;
(ii) is $\delta$-closed (respectively $\delta^{free}$-closed);
and (iii) has no component of $K$-degree higher than $k$, then
$a = \delta b$ (respectively, $a = \delta^{free} b$), where $b$
has also no component of $K$-degree higher than $k$.

One easily verifies that
\begin{eqnarray}
[ K, \partial_\mu ] =\partial_\mu.
\end{eqnarray}
Furthermore, if $a$ has $K$-degree
$k$, then, $\delta a/\delta A^a_\mu$ and $\delta a/\delta y^i$
have $K$-degree $k$, $\delta a/\delta y^*_i$ has $K$-degree $k-1$,
$\delta a/\delta A^{*\mu}_a$ has $K$-degree
$k-2$, while $\delta a/\delta C^*_a$ has $K$-degree $k-3$.
Finally, any $q$-form with bounded $K$-degree is necessarily a polynomial
in the derivatives of the fields, the antifields and their derivatives,
since these variables have all strictly positive $K$-degree.  It may,
however, be an infinite series in the undifferentiated fields, which
carry zero $K$-degree.

The existence of a $K$-degree with the above properties
is the characteristic feature of the so-called ``normal
theories".  This concept applies to reducible or irreducible
gauge theories and is captured as follows.
Let ${\cal L}(\phi,\partial\phi,\partial^2\phi,
\dots,\partial^s\phi)$ be the Lagrangian of a theory,
\begin{eqnarray}
{\cal L}={\cal L}^{free}+{\cal L}^{int}.\label{north}
\end{eqnarray}
The free term is quadratic in the fields and their derivatives.
We shall say that (\ref{north}) describes a
normal theory iff

(i) the free theory and the full theory have the same number of gauge
invariances, with the same reducibility properties,
so that $\delta^{free}$
is acyclic at positive antighost number (with the antifield spectrum
of the full theory);

(ii) it is possible to define an even derivation $K$ along
the lines of the Yang-Mills case, which is the sum
of the operator counting the derivatives and an operator
$A$ commuting with $\partial_\mu$,
$K=N_\partial + A$.
The even derivation $ A$ should assign strictly positive degree
to the antifields, and non negative degree to the fields $\phi^i$.
The even derivation $K$ should be
such that the differential $\delta^{free}$ has only components
of non positive $K$-degree
\begin{eqnarray}
\delta^{free}=\sum_t \delta^{free,t},\qquad
[K,\delta^{free,t}]=t\delta^{free,t},\qquad
t\leq 0 ,
\end{eqnarray}
and
\begin{eqnarray}
[ K, \partial_\mu ] =\partial_\mu.
\end{eqnarray}
Furthermore, the zeroth order differential $\delta^{free,0}$
should be acyclic at positive antighost number~; as
in the Yang-Mills case,
the $A$-weight of the antifields $\phi^*_i$ is determined by the
differential order of the corresponding free equations of motion~;
the $A$-weight of the antifields $C^*_\alpha$ (and
$C^*_\Delta$ if any) is determined by the differential
order of the corresponding reducibility identities and the
$A$-weight of the previous antifields $\phi^*_i$ (or
$C^*_\alpha$)~;

(iii) finally, the interacting part of $\delta$
must contain only terms of non positive $K$-degree,
\begin{eqnarray}
\delta^{int}=\sum_t \delta^{int,t},\qquad
[K,\delta^{int,t}]=t\delta^{int,t},\qquad
t\leq 0 .
\end{eqnarray}
This condition expresses that there are at most as many derivatives in
$\delta^{int}$ as there are in $\delta^{free}$.  Note that we do not require
$\delta^{int,0}$ to vanish, but that the sum $\delta^0 = \delta^{free,0}
+ \delta^{int,0}$ is always acyclic because of (ii).

It may happen that condition (ii) is fulfilled only after one has
redefined the fields.
Einstein gravity (with or without cosmological constant) is a normal theory,
characterized by a non-vanishing $\delta^{int,0}$.  Even though
one can split the derivatives as in section 3, a theory that is not a normal
theory is
\begin{eqnarray}
{\cal L}=\phi\qed^2\phi+(\partial_\mu\phi\partial^\mu\phi)^{10}
\label{badnorm}
\end{eqnarray}
since the interaction vertices contain 20 derivatives while the free part
contains only 4 derivatives.

Let $a$ be a solution of
$\delta a + db =0$, with antighost number $\geq 3$ and bounded
$K$-degree. Then, $a$ is a polynomial in the antifields, their
derivatives and the derivatives of the fields, with coefficients
that may be infinite series in the undifferentiated fields.
Let us expand $a$ according to its polynomial degree,
$a = a_2 + a_3 + a_4 + ...$.
The lower index denotes the polynomial degree of $a$
(not the $K$-degree) and the series terminates if
$a$ is polynomial in the undifferentiated fields.
The first term is at least quadratic because we assume
the antighost number of $a$ to be $\geq 3$.
The term of degree
$2$ in the cocyle condition reads
\begin{eqnarray}
\delta^{free} a_2 + db_2 =0.
\end{eqnarray}
By theorem \ref{9.1}, this implies
\begin{eqnarray}
a_2=\delta^{free} c_2+de_2.
\end{eqnarray}
Thus, $a-\delta c_2 - de_2$ has no quadratic piece and reads
$a-\delta c_2 - de_2=a^\prime_3+a^\prime_4+\dots $. That is,
one can remove $a_2$ through the addition of $\delta$-exact modulo $d$
terms. One can repeat the argument to remove successively $a_3$,
$a_4,\dots$. This shows that $a$ is $\delta$-trivial modulo $d$,
$a=\delta c + de$. The conclusion is correct, however, only if one can
prove that the procedure does not introduce arbitrarily high
derivatives of the variables.
The question is not entirely straightforward because when one removes
$a_i$, one generically modifies the next terms $a_{i+1}$ and $a_{i+2}$.
Thus, even if $a_{i+1}=a_{i+2}=0$ originally, one may have
$a_{i+1}\neq 0$ and $a_{i+2}\neq 0$ after $a_i$ has been set equal to
zero by the addition of a $\delta$-exact modulo $d$ term.

It is here that the fact that ${\cal L}^{free}$ contains the maximum number of
derivatives, or more precisely, that the theory is a normal theory,
plays a crucial role. Indeed, the components of
$a$ are bounded in $K$-degree, let us say by $k$.
Then the reconstruction formula
(\ref{9.14}) and (\ref{9.10})-(\ref{9.12})
show that the $K$-degree of $c_2$, given by the first
terms in the right-hand side of (\ref{9.14}), cannot exceed
$k$. It then follows that the $K$-degree of $e_2$ cannot
exceed $k-1$ since $[K,\partial_\mu ] = \partial_\mu$.
Therefore, the term $\delta^{int}c_2$, which
modifies $a_3$, $a_4$, etc has $K$-degree smaller than
(or equal to if $\delta^{int,0}$ does not vanish)
$k$. The same reasoning applies next to $c_3$, $e_3$, $c_4$,
$e_4,\dots$. We can thus conclude
that the $K$-degree of $c$, respectively $e$, does not exceed
$k$, respectively $k -1$.  Thus, $c$ and $e$ are polynomial in the
derivatives of the fields, the antifields, and their derivatives.
Moreover, if $\delta^{int,0}$ is absent and if the initial $a$ is
a polynomial of order $L$ in all the variables
and their derivatives, $a = a_2 + a_3 + \dots + a_L$, then
the process of successively eliminating $a_2$, $a_3,\dots$
stops after at most $L+k$ steps.
This is because the $K$-degree strictly decreases
at each step (it cannot remain
equal to $k$). Accordingly, $c$ and $e$ are polynomial
not just in the derivatives, but also in the undifferentiated variables.

We have thus established:
\begin{theorem}\label{normal1}{\em{\bf :}}
let ${\cal L}$ be the Lagrangian of a normal, reducible gauge theory of order
$r$.
Then
\begin{eqnarray}
H_k(\delta|d)=0\label{stop}
\end{eqnarray}
for $k>r+2$ in the space of forms with coefficients that are polynomials in
the differentiated variables and the antifields, and formal series in the
undifferentiated fields.
\end{theorem}

\begin{theorem}\label{normal2}{\em{\bf :}}
If, in addition, $\delta^{int,0} = 0$, then
\begin{eqnarray}
H_k(\delta|d)=0\label{stop}
\end{eqnarray}
for $k>r+2$ in the space of forms with coefficients that are polynomials in
all the variables and their derivatives.
\end{theorem}

Theorem 10.1 applies to gravity, where the infinite series
$g^{\mu \nu} = \eta^{\mu \nu} - h^{\mu \nu} + \dots $ are allowed;
while Theorem 10.2 applies to Yang-Mills theory.

Since reducible gauge theories of order $r$ are usually not only
normal theories, but also theories of
Cauchy order $r+2$, Theorems 10.1 (or 10.2) and
8.3 are equivalent in practice.  However, the conditions under which
they apply may in principle be different (see (\ref{badnorm}) and section
14 below on shift symmetry) and so, Theorems 10.1 (or 10.2)
and 8.3 are in
those cases inequivalent.

Finally, we point out that it would be of interest to extend the
conditions under which the perturbative argument behind
Theorems 10.1 and 10.2 applies.

\section{Results on $H^n_2(\delta|d)$}
\setcounter{equation}{0}
\setcounter{theorem}{0}

It follows from the above analysis that
$H_j(\delta|d)$ vanishes whenever $j>2$
for the usual
irreducible gauge theories.
We shall now establish some
general theorems on $H^n_2(\delta|d)$.

Let $a$ be a representative of $H^n_2(\delta|d)$. By adding a
$\delta$-exact term and a total
derivative if necessary, one has
\begin{eqnarray}
a=f^\alpha C^*_\alpha +\mu
\end{eqnarray}
where $f^\alpha$ may be assumed to depend on $x_A$ only and
where $\mu$ is quadratic in the antifields
$\phi^*_i$ and their derivatives.

\begin{theorem}\label{var}{\em{\bf :}}
a necessary condition for $a$ to be a $\delta$-cycle modulo $d$ is
that $f^\alpha$ be the
parameter of a gauge transformation
that leaves the fields invariant on-shell,
\begin{eqnarray}
f^\alpha R^i_\alpha+
\partial_\mu f^\alpha R^{i\mu}_\alpha+
\dots+\partial_{\mu_1\dots\mu_k}f^\alpha
R^{i\mu_1\dots\mu_k}_\alpha\approx 0\label{glob}
\end{eqnarray}
(``global reducibility identity").
\end{theorem}

\proof{the proof is direct. One has
\begin{eqnarray}
\delta a=(f^\alpha R^i_\alpha+
\partial_\mu f^\alpha R^{i\mu}_\alpha+
\dots+\partial_{\mu_1\dots\mu_k}f^\alpha
R^{i\mu_1\dots\mu_k}_\alpha)\phi^*_i+
\delta\mu+\partial_\mu k^\mu
\end{eqnarray}
where $\delta\mu$ vanishes on-shell. The Euler-Lagrange derivative
of $\delta a + db =0$ with respect
to $\phi^*_i$ yields then (\ref{glob}), as desired.}

Thus, if there is no solution $f^\alpha$ to (\ref{glob}), one may
assume that $a$ is quadratic
in the antifields $\phi^*_i$ and their derivatives. This occurs in
electromagnetism with
charged matter fields since then (\ref{glob}) reads
\begin{eqnarray}
\partial_\mu f\approx 0,\qquad ief\psi\approx 0
\end{eqnarray}
from which it follows that $f\approx 0$ ($e\neq 0$) and thus $f=0$
($f$ depends only on $x_A$).
This also occurs in (i) Yang-Mills theory with a semi-simple gauge
group, for which (\ref{glob})
becomes
\begin{eqnarray}
D_\mu f^a\equiv\partial_\mu f^a-gC^a_{bc}A^b_\mu f^c\approx 0
\end{eqnarray}
which has no solution $f^a(A^a_\mu,\partial_\rho A^a_\mu,\dots)$
besides $f^a = 0$~;
and (ii) Einstein gravity, for which
(\ref{glob}) reads
\begin{eqnarray}
\xi_{\alpha;\beta}+\xi_{\beta;\alpha}\approx 0
\end{eqnarray}
which again has no solution $\xi_\alpha(g_{\rho\sigma},\partial_\lambda
g_{\rho\sigma},\dots)$ besides $\xi_\alpha = 0$ (a generic metric has no
Killing vectors~; thus
$\xi_\alpha(g_{\rho\sigma},\partial_\lambda
g_{\rho\sigma},\partial_{\lambda\mu}g_{\rho\sigma},\dots)
=0$ for generic $g_{\rho\sigma}$'s and by continuity, $\xi_\alpha
=0$).

Consider now a solution $\mu$ of $\delta\mu+\partial_\mu b^\mu
=0$ which is purely quadratic in
the antifields and their derivatives.

\begin{theorem}\label{quadr}{\em{\bf :}}
for linear gauge theories, there is no nontrivial element of
$H^n_2(\delta|d)$ that is purely
quadratic in the antifields $\phi^*_i$ and their derivatives.
That is, if $\mu$ is quadratic in
the antifields $\phi^*_i$ and their derivatives and if
$\delta\mu +\partial_\mu b^\mu =0$ then
$\mu =\delta C+\partial_\mu V^\mu$.
\end{theorem}

\proof{the proof proceeds as the proof of theorem \ref{9.1}. One computes
first the variational derivatives of $\mu$ with respect to $\phi^*_i$ and
$\phi^i$ from the cycle condition $\delta\mu
+\partial_\mu b^\mu =0$.
One then reconstructs $\mu$ by a formula analogous to (\ref{9.14})
recalling that $\delta\mu /\delta C^*_\alpha =0$ since $\mu$ is purely
quadratic in the antifields $\phi^*_i$ and their derivatives. This yields
immediately the desired result.}

Again, theorem \ref{quadr} can be extended to non linear,
normal theories, as the perturbative
argument of the previous section indicates.

\section{Calculation of $H^n_2(\delta|d)$ for electromagnetism}
\setcounter{equation}{0}
\setcounter{theorem}{0}

As we have seen, electromagnetism and Yang-Mills models have Cauchy order 2.
Hence, only
$H^n_2(\delta|d)\simeq H^{n-1}_1(\delta|d)$ can be different from
zero besides
$H^n_1(\delta|d)$. The explicit calculation of $H^n_1(\delta|d)$ is a
difficult question that depends
explicitly on the model under consideration and its rigid symmetries.
It turns out that, by contrast,
the calculation of $H^n_2(\delta|d)$ can be carried out completely.
We first compute
$H^n_2(\delta|d)$ for free electromagnetism. The Koszul-Tate
differential acting on the undifferentiated
generators reads explicitly
\begin{eqnarray}
\delta A_\mu =\delta C =0,\ \delta A^{*\mu}=-\partial_\rho
F^{\rho\mu},\ \delta C^*=-\partial_\mu
A^{*\mu}.
\end{eqnarray}
\begin{theorem}\label{2_em}{\em{\bf :}}
For a free abelian gauge field $A_\mu$, the groups $H^n_2(\delta|d)$
and $H^{n-1}_1(\delta|d)$
are one-dimensional. One can take as representatives:
\begin{eqnarray}
for\ H^n_2(\delta|d)\ : C^* dx^0\wedge\dots\wedge dx^{n-
1}\label{s_1}\\
for\ H^{n-1}_1(\delta|d)\ : {1\over (n-
1)!}A^{*\mu}\varepsilon_{\mu\alpha_1\dots\alpha_{n-1}}
dx^{\alpha_1}\wedge\dots\wedge dx^{\alpha_{n-1}}\label{s_2}
\end{eqnarray}
(or, in dual notations, $C^*$ and $A^{*\mu}$, respectively).
\end{theorem}

\proof{we consider explicitly $H^n_2(\delta|d)$ and work in dual
notations. By adding
a divergence if necessary, any representative $a$ of an element of
$H^n_2(\delta|d)$
can be chosen to be of the form $a=C^*f(x_A)+\mu$ where $\mu$
depends on the fields
and is quadratic in the antifields $A^{*\mu}$. One has $\delta a
+\partial_\mu b^\mu =0$ for some
$b^\mu$. By theorem \ref{var}, the function $f$ must fulfill
$\partial_\mu f\approx 0$, i.e.,
$df +\delta k=0$ for
some $k$ of antighost number 1 and form degree 1. But $
H^0_0(d|\delta)/{\bf R}\simeq H^1_1(\delta|d)$
vanishes (we assume the spacetime dimension to be strictly greater than
2). Accordingly,
$f$ is trivial, $f\approx const$. Since $f$ does not involve the
equations of motion, this
forces $f=const.$ (strongly).

The cycle $fC^*$ with $f=const.$ is a solution of
$\delta (fC^*) +\partial_\mu (fA^{*\mu})=0$
by itself. By substracting it from $a$, one may assume $a=\mu$
where $\mu$ is quadratic in the
antifields $A^{*\mu}$. Using theorem \ref{quadr}, one then finds that
$\mu$ is
$\delta$-trivial modulo $d$ and does not contribute to the
cohomology.
This completes the demonstration of the theorem.}

Remarks: (i) The solutions (\ref{s_1}) and (\ref{s_2}) are non trivial
because they do not contain
derivatives of the fields (while a trivial term $\delta m +dn$ contains
necessarily derivatives).

(ii) The condition $n>2$ is essential. For $n=2$ there are other non-
trivial cocycles. For instance,
$a={1\over 2}\varepsilon_{\mu\nu}F^{\mu\nu}C^*+{1\over
2}\varepsilon_{\mu\nu}A^{*\mu}A^{*\nu}$
fulfills $\delta a + db =0$ and is non trivial because it contains
$F^{\mu\nu}$, $C^*$ and
$A^{*\mu}$ undifferentiated. To analyze completely the 2-
dimensional case, we use the
chain of isomorphisms $H^2_2(\delta|d)\simeq H^1_1(\delta|d)\simeq
H^0_0(d|\delta)/{\bf R}$. Thus it is
enough to find all non-trivial 0-forms $f$ that are closed (constant)
modulo the equations of motion.
These 0-forms may be assumed to depend on the $x_A$ only. It is
convenient at this stage to
adopt a different parametrisation of the field variables: one may
express any function of the $A_\mu$
and their derivatives in terms of $A_\mu$, its symmetrized
derivatives $A_{\mu_1\dots\mu_k}=
\partial_{(\mu_1}\dots\partial_{\mu_{k-1}}A_{\mu_k)}$ (which are
independent), $F_{01}\equiv
{1\over 2}\varepsilon_{\alpha\beta}F^{\alpha\beta}\equiv ^*F$ and
its derivatives
(which are also independent, the Bianchi identities are empty in two
dimensions). The equations
of motion set the derivatives of $^*F$ equal to zero and leave
$A_\mu$ and its symmetrized derivatives
free. We may assume
$f=f(A_\mu,A_{\mu\nu},\dots,A_{\mu_1\dots\mu_k},F_{01})$ since $\partial_\rho
F_{01}\approx 0$.
Impose now
$\partial_\rho f\approx 0$. One has $\partial_\rho f =(\partial
f/\partial A_\mu)
\partial_\rho A_\mu +\dots +(\partial f/\partial
A_{\mu_1\dots\mu_k})
\partial_\rho A_{\mu_1\dots\mu_k} +(\partial f/\partial F_{01})
\partial_\rho F_{01}$. The last term is weakly zero. Since
$\partial_\rho A_{\mu_1\dots\mu_k}\approx
{1\over k+1}A_{\rho\mu_1\dots\mu_k}$ and since
$A_{\rho\mu_1\dots\mu_k}$ is unconstrained by the equations of
motion,
the requirement $\partial_\rho f\approx 0$ imposes $\partial
f/\partial A_{\mu_1\dots\mu_k}
\approx 0$, i.e., $\partial f/\partial A_{\mu_1\dots\mu_k}=0$ ($f$
does not involve the
derivatives of $F_{01}$). Thus, $f$ cannot depend on the
symmetrized derivatives of order $k$.
Similiarily, it cannot depend on the symmetrized derivatives of order
$k-1$, etc \dots, i.e.,
$f=f(F_{01})$. Any function of $F_{01}$ is a solution of the problem
and is non trivial.
Accordingly, $ H^0_0(d|\delta)\simeq C^\infty(F_{01})$. Note that
$H^0_0(d|\delta)\simeq
H^0(d,H_0(\delta))$ and is an algebra.

\section{Calculation of $H^n_2(\delta|d)$ for Yang-Mills
models and gravity}\label{delta_d}
\setcounter{equation}{0}
\setcounter{theorem}{0}
The previous section shows that $H^n_2(\delta|d)$ is non empty for
free electromagnetism,
because there is then a
global reducibility identity on the gauge
transformations\footnote{There is no contradiction between the fact
that electromagnetism is a irreducible
gauge theory because gauge transformations should vanish at
infinity. This kills constant gauge
parameters. However, in analyzing $H(\delta|d)$, no boundary
condition is imposed.}.
Now, there is no global reducibility identity when self-couplings or
couplings to matter
are included. Indeed, gauge transformations leaving the Yang-Mills
field $A^a_\mu$
and the matter fields $y^i$ invariant should fulfill
\begin{eqnarray}
D_\mu f^a \approx 0,\ f^a (T_a)^i_j y^j\approx 0\label{glob_red},
\end{eqnarray}
whose only solution $f^a(A^a_\mu,\partial_\rho A^a_\mu,\dots,y^i,\partial_\mu
y^i,\dots)$ is $f^a \approx
0$, and thus $f^a=0$ if one assumes - as one can -
that $f^a$ involves only the $x_A$.
Accordingly, by theorem \ref{var}, an element of $H^n_2(\delta|d)$
should be quadratic in the antifields of antighost
number 1. Since the Yang-Mills theory is a normal theory,
theorem \ref{quadr} implies then that
$H^n_2(\delta|d)$ is empty. Non-vanishing cohomology arises only when there
are uncoupled abelian factors since equation (\ref{glob_red}) has then
non trivial solutions.

We have thus proved

\begin{theorem}{\em{\bf :}}
In spacetime dimensions $\geq 3$, the group $H^n_2(\delta|d)$
vanishes unless there are
free abelian gauge fields $A^\alpha_\mu$. In that case, a basis of
$\delta$-cycles modulo $d$
is given by
\begin{eqnarray}
f^\alpha C^*_\alpha,\ f^\alpha=const.
\end{eqnarray}
where the $C^*_\alpha$ are the antifields of antighost number 2
associated with the uncoupled abelian
factors.
\end{theorem}

The same reasoning applies also to Einstein gravity: linearized gravity
has ten global reducibility identities corresponding to the ten Killing
vectors of Minkowski space. These define cohomological
classes of $H_2^n(\delta|d)$.
The full theory, however, has no global reducibility identity (a
generic solution of Einstein equations has no Killing vector). Thus
$H_2^n(\delta|d)$ is empty in Einstein gravity.

Comments: (i) It follows from $H^n_2(\delta|d)=0$ and the
isomorphism $H^n_2(\delta|d)\simeq
H^{n-2}_0(d|\delta)$ that there is no local $(n-2)$-form that is closed
modulo the
equations of motion for generic Yang-Mills models, or Einstein gravity.

(ii) In two dimensions, one may compute $H^2_2(\delta|d)\simeq
H^0_0(d|\delta)/{\bf R}$ directly.
The analysis proceeds as in the Abelian case. The equations of motion
are equivalent to
$D_\mu F^a_{01}=0$ (in the absence of coupling to matter) and thus,
eliminate the derivatives of
$F^a_{01}$. Any function
$f(A^a_\mu,A^a_{\mu\nu},\dots,A^a_{\mu_1\dots\mu_k},F^a_{01})$
solution of $df+\delta m=0$ must be gauge invariant since $\gamma
df+\gamma\delta m =0$
implies $\gamma m=\delta u + dv$ and hence $\gamma f -\delta
v=0$, i.e., $\gamma f =0$ ($f$
does not contain the derivatives of $F^a_{01}$). This means that $f$
must be an invariant
function of $F^a_{01}$. Any such function fulfills $df\approx 0$ and
is thus a solution.
That is, $H^0_0(d|\delta)$ is isomorphic to the set of invariant
functions of $F^a_{01}$ in the
absence of matter. When couplings to matter are included, however,
$H^0_0(d|\delta)/{\bf R}$
is generically zero dimensional.

\section{Non minimal sector - Shift symmetry}
\setcounter{equation}{0}
\setcounter{theorem}{0}
In order to fix the gauge, it is often useful to add further variables
known as the
``variables of the non-minimal sector". This procedure is physically
acceptable because these
variables do not modify the BRST cohomology $H^*(s)$
\cite{Henneaux2}. We show that they do not modify the
local BRST cohomology $H^*(s|d)$ either.

The standard non minimal sector contains the ``antighosts" $\bar
C_a$, the auxiliary fields $b_a$
and the corresponding antifields $\bar C^{*a}$ and $b^{*a}$. The
action of the BRST differential
on those variables is
\begin{eqnarray}
s\bar C_a =b_a,\ sb_a =0,\ s\bar C^{*a}=0,\ sb^{*a}=-\bar
C^{*a}\label{nonmin}.
\end{eqnarray}
This is the caracteristic form of a contractible differential algebra. As
usual, one extends $s$
to the derivatives of $\bar C_a$, $b_a$, $\bar C^{*a}$ and $b^{*a}$ so
that
$s\partial_\mu=\partial_\mu s$.

The triviality of $s$ in the non minimal sector is proved
by introducing
the contracting homotopy $\rho$,
\begin{eqnarray}
\rho =\sum_{k\geq 0}[{\partial^R\over\partial
(\partial_{\mu_1\dots\mu_k}b_a)}\partial_{\mu_1\dots\mu_k} \bar
C_a-{\partial^R\over\partial
(\partial_{\mu_1\dots\mu_k}\bar C^{*a})}\partial_{\mu_1\dots\mu_k}
b^{*a}]
\end{eqnarray}
such that $\rho s + s\rho=\tilde N$, where $\tilde N$ counts the
number of variables of
the non minimal sector and their derivatives. The crucial feature of
the contracting homotopy
$\rho$ is that it commutes with $\partial_\mu$,
$\rho\partial_\mu=\partial_\mu\rho$.
Thus, it anticommutes with $d$,
\begin{eqnarray}
\rho d + d\rho =0\label{acomm}.
\end{eqnarray}
The existence of a contracting homotopy that commutes with
$\partial_\mu$ follows from the fact that
the action of $s$ on the fields of the non minimal sector does not
increase the order
of the derivatives. Such a homotopy does not exist for $\delta$ in the
minimal sector whenever $H_k(\delta|d)$ is non
trivial \cite{Henneaux3}.

Because of (\ref{acomm}), one may easily establish that the non
minimal sector does not contribute to
$H(s|d)$. Let $a$ be a solution of $sa+db=0$. Decompose $a$ according
to the its $\tilde N$-degree,
$a=\Sigma_{k\geq 0}a_k$. The term $a_0$ does not contain the
variables of the non-minimal sector.
{}From $sa+db=0$, one infers
\begin{eqnarray}
a-a_0=\Sigma_{k\geq 1}{\tilde N a_k\over k}=\Sigma_{k\geq
1}(\rho s + s\rho){a_k\over k}
=s(\Sigma_{k\geq 1}\rho{a_k\over k})+d(\Sigma_{k\geq
1}\rho{b_k\over k}).
\end{eqnarray}
Hence, $a-a_0$ is $s$-exact modulo $d$ and can be removed from
$H(s|d)$.

We have thus proved
\begin{theorem}{\em{\bf :}}
The variables of the non minimal sector do not contribute to $H(s|d)$:
one can remove
the variables of the non minimal sector from any $s$-cocycle modulo
$d$ by adding to it
a $s$-boundary modulo $d$.
\end{theorem}

In particular, in Yang-Mills theory, one may analyze the local BRST
cohomology in terms of the
original variables of the ``minimal sector" $A^a_\mu$, $C^a$,
$A^{*\mu}_a$, $C^*_a$, $y^i$, $y^*_i$
introduced above.

A similar analysis applies to gauge symmetries not involving the
derivatives of the fields,
such as the ``shift symmetry" of \cite{Alfaro}. Consider a gauge
theory such that (i) the
fields $\varphi^i$ split into two groups $\varphi^i\equiv
(e^a,\omega^\alpha)$~; and (ii)
the gauge transformations are mere translations in the
$\omega^\alpha$,
\begin{eqnarray}
\delta_\varepsilon e^a=0\label{shift1}\\
\delta_\varepsilon\omega^\alpha=\varepsilon^\alpha\label{shift2}.
\end{eqnarray}
The Lagrangian depends only on the $e^a$ and their derivatives. We
shall call (\ref{shift1}),
(\ref{shift2}) ``shift symmetry" because (\ref{shift2})
is just a shift of the
$\omega$'s. An example (but not the only one) of such a gauge
theory is obtained by replacing the fields $\phi^i$ by $\phi^i-\psi^i$
in a theory without
gauge invariance. The Lagrangian is then invariant under the shifts
$\delta_\varepsilon \phi^i=
\varepsilon^i$, $\delta_\varepsilon \psi^i=\varepsilon^i$. The
redefinition $e^i={1\over 2}
(\phi^i-\psi^i)$, $\omega^i=(\phi^i+\psi^i)$ brings the theory to the
desired form
(\ref{shift1}), (\ref{shift2}). Since this redefinition is invertible and
local, it does not
modify $H^*(s|d)$. For instance, if one starts with the Klein-Gordon theory,
one gets the Lagrangian ${\cal L}(e,\omega,\partial e,\partial \omega)
\equiv -{1\over 2} \partial_\mu e\partial^\mu e$. The theory has
Cauchy order one (the $x_A$ are $e$, $\partial_\rho e$ and their
spatial derivatives together with $\omega$ and its derivatives). It is
also a linear irreducible gauge theory. Thus, by theorem \ref{9.1},
$H_j(\delta|d)=0$ for $j>2$. This is strengthned by theorem
\ref{rest_norm_theo} to $H_j(\delta|d)=0$ for $j>1$ (Cauchy order 1).

The BRST cohomology of the shift symmetry can be completely
computed. Indeed, the BRST transformation for
(\ref{shift1}), (\ref{shift2}) reads
\begin{eqnarray}
se^a =0,\ se^*_a =-{\delta {\cal L}_0\over \delta e^a}\\
s\omega^\alpha =C^\alpha,\ sC^\alpha =0,\ s\omega^*_\alpha =0,\
sC^*_\alpha=\omega^*_\alpha
\label{shift3}.
\end{eqnarray}
The transformation (\ref{shift3}) takes the same form as
(\ref{nonmin}). Thus, the same argument shows
that $\omega^\alpha$, $C^\alpha$, $\omega^*_\alpha$ and
$C^*_\alpha$ do not contribute to the
cohomology $H^*(s|d)$. Only the gauge invariant degrees of freedom
$e^a$ and $e^*_a$ contribute
to $H^*(s|d)$. In particular, one has
\begin{theorem}{\em{\bf :}}
The shift symmetry cannot be anomalous, $H^{1,n}(s|d)=0$.
\end{theorem}

These results can be straightforwardly extended to the case of a
gauge group that is the direct
product of a shift symmetry group by another group. One may
always reshuffle terms in
$H^{1,n}(s|d)$ so that the shift symmetry remains anomaly-free
\cite{DeJonghe}.

\section{Auxiliary fields}
\setcounter{equation}{0}
\setcounter{theorem}{0}

The cohomological groups $H(s|d)$ and $H(\delta|d)$ are manifestly
invariant under invertible, local
change of variables. We shall now show that they are also invariant
under the introduction of
so called ``auxiliary fields".

If the fields $\phi^i$ split as $(\phi^i) =
(y^{\bar\alpha},z^\alpha)$
where the $z^\alpha$ are such that the equations of motion $\delta
S_0/\delta z^\alpha =0$
can be solved for $z$,
\begin{equation}
{\delta S_0 \over \delta z^\alpha}=0 \Longleftrightarrow z^\alpha =
Z^\alpha
(y^{\bar\alpha},\partial_\mu
y^{\bar\alpha},\dots,\partial_{\mu_1\dots\mu_k} y^{\bar\alpha}),
\label{aux}
\end{equation}
where $Z^\alpha$ are local functions, one says that the $z^\alpha$ are
``auxiliary fields".
Given a theory with auxiliary fields, one defines the reduced action
$\bar {S}_0 [y]$ by eliminating the auxiliary fields from $S_0[y,z]$
using their own
equations of motion
\begin{equation}
\bar {S}_0 [y]=S_0[y,z=Z(y)] .
\end{equation}
The theories based on $S_0[y,z]$ and $\bar {S}_0 [y]$ are classically
equivalent.
They are also quantum-mechanically equivalent, at least formally
\cite{Henneaux5}.

Auxiliary fields can be useful for closing gauge algebras off-shell
and occur at various places in physics.
The conjugate momenta of the Hamiltonian formalism can be viewed as
auxiliary fields.
Other examples of auxiliary fields are the field strengths in the first
order formulation of
Yang-Mills theory,
\begin{eqnarray}
S_0[A_\mu^a,H_{\mu\nu}^a]=\int d^nx\ H^a_{\mu\nu} H_a^{\mu\nu}
+H^a_{\mu\nu}(\partial^\mu A^\nu_a-
\partial^\nu A^\mu_a+f_{abc}A^{\mu b}A^{\nu c}).
\end{eqnarray}
In gravity, the Christoffel coefficients
$\Gamma^\alpha_{\beta\gamma}$ in the Palatini formulation of
the Hilbert action are auxiliary fields, as are the connexion
components $\omega_{ab\mu}$ in
the first order tetrad formalism.

For the subsequent discussion, it is useful to define $T[y,z]$ through
\begin{eqnarray}
S_0[y,z]=\bar S_0 [y]+T[y,z]\label{11.4}.
\end{eqnarray}
The equations $\delta S_0/\delta z=0$ coincide with $\delta T/\delta
z=0$ and one has \cite{Henneaux5}
\begin{eqnarray}
{\delta T\over\delta y^i(x)}=\int \mu^A_i(x,x^\prime){\delta T\over
z^A(x^\prime)}dx^\prime\label{11.5}
\end{eqnarray}
where $\mu^A_i(x,x^\prime)$ is a combination of
$\delta(x,x^\prime)$ and its derivatives.

The relationship between the BRST cohomologies of two formulations
of the same theory
differing in auxiliary field content is easily derived by following the
approach of
\cite{Henneaux5}. As shown in \cite{Henneaux5}, one may redefine
the gauge transformations
in such a way that the gauge transformations $\delta_\varepsilon
y^i$ of the theory
with auxiliary fields coincide with the gauge transformations of the
reduced theory and
\begin{eqnarray}
\delta_\varepsilon z^A(x)=-\int
\mu^A_i(x,x^\prime)\delta_\varepsilon y^i(x^\prime)dx^\prime.
\end{eqnarray}
With that choice, a solution $S$ of the master equation of the full
theory is given by
\begin{eqnarray}
S=\stackrel{(0)}{S}+\stackrel{(1)}{S}+\sum_{k\geq
2}\stackrel{(k)}{S}
\end{eqnarray}
where $\stackrel{(0)}{S}$ is a solution $\bar S$ of the master
equation of the reduced theory and
where $\stackrel{(1)}{S}$ is given by
\begin{eqnarray}
\stackrel{(1)}{S}=-\int z^*_A(x^\prime) \mu^A_i(x^\prime,x){\delta
\stackrel{(0)}{S}\over\delta y^*_i(x)}dx dx^\prime.
\end{eqnarray}
The index $k$ denotes the number of antifields $z^*_A$ in
$\stackrel{(k)}{S}$ (not the antighost
number) and the terms $\stackrel{(k)}{S}$ ($k\geq 2$) are
successively determined by the method of homological perturbation
theory by equations of the form
\begin{eqnarray}
\delta^\prime
\stackrel{(k+1)}{S}=\stackrel{(k)}{D}(\stackrel{(0)}{S},\dots,\stackrel{
(k)}{S}),\qquad
k=1,2,3,\dots\label{11.9}
\end{eqnarray}
where (i) $\stackrel{(k)}{D}$ involves the antibracket of the $S$'s of
lower order~; and
(ii) $\delta^\prime$ (acting from the left like in \cite{Henneaux5})
is the Koszul-Tate resolution
of the surface
where the auxiliary fields are on-shell,
\begin{eqnarray}
\delta^\prime z^*_A={\delta S_0\over \delta z^A},\qquad
\delta^\prime(everything\ else)=0\label{zeq}.
\end{eqnarray}

Because the equations of motion (\ref{zeq}) for $z^A$ are equivalent
to algebraic equations
$z^A=Z^A(y,\partial y,\dots)$, the theory based on the equations
(\ref{zeq}), viewed as equations
for $z^A$ with fixed $y$'s, is a normal theory of order 0. Indeed,
there is no independent
derivatives of $z^A$ since they are all determined by the equations
(\ref{zeq}) and their
derivatives. Thus the set $I_0$ (for the $z$'s) is empty and
$\partial_\alpha I_0$ is
clearly contained in $I_0$. By theorem \ref{rest_norm_theo}, one
concludes that $H_k(\delta^\prime|d)
=0$ for $k>0$ (besides $H_k(\delta^\prime)=0$).

It follows from the standard method of homological perturbation that
the terms
$ \stackrel{(k+1)}{S}$ subject to (\ref{11.9}) exist and can be taken to
be local functionals.
Similarily, let $\bar A[y,y^*]$ be a local functional solution of $(\bar
A,\bar S)=0$.
Then there exist a functional $A$,
\begin{eqnarray}
A=\bar A+\stackrel{(1)}{A}+\Sigma_{k\geq 2}\stackrel{(k)}{A}\\
\stackrel{(1)}{A}=-\int z^*_A(x^\prime) \mu^A_i(x^\prime,x){\delta
\bar A\over\delta y^*_i(x)}dx dx^\prime.
\label{11.11}
\end{eqnarray}
which solves $(A,S)=0$. This functional is determined recursively by
equations of the same form as
(\ref{11.9}),
\begin{eqnarray}
\delta^\prime
\stackrel{(k+1)}{A}=\stackrel{(k)}{F}(\stackrel{(0)}{A},\dots,\stackrel
{(k)}{A})
\end{eqnarray}
where $\stackrel{(k)}{F}$ involves the brackets of the lower order
$\stackrel{(i)}{A}$'s with the
$\stackrel{(j)}{S}$. Again, each term in the expansion (\ref{11.11}) is
a local functional because
$H_k(\delta^\prime|d)=0$. We have thus proved

\begin{theorem} {\em{\bf :}}\label{11.1}
The BRST cohomology groups $H(s|d)$ and $H(\bar s|d)$ respectively
associated
with two different formulations of the same theory differing only in
the auxiliary field content
are isomorphic.
\end{theorem}

This theorem is the analog for local functionals of the isomorphism
theorem
$H(s)\simeq H(\bar s)$ that holds for local $p$-forms or arbitrary
functionals.
It can easily be extended to the generalized auxiliary fields
introduced in \cite{Dresse}
(see also \cite{Brandt2}) as we now show.

Assume that the solution of the master equation $S[y,y^*,z,z^*]$ is such
that the equations
${\delta S/\delta z^A=0}$ can be solved at $z^*_A=0$ for the $z^A$ as
functions of the
$y^i$ {\it and} $y^*_i$,
\begin{eqnarray}
{\delta S\over\delta z^A}\mid_{z^*=0}=0\Longleftrightarrow
z^A=Z^A(y,\partial y,\dots,y^*,\partial y^*,\dots)
\label{11.14}
\end{eqnarray}
If this is the case, one says that the $z^A$ are generalized auxiliary
fields. Ordinary auxiliary
fields are a particular case of (\ref{11.14})~; they do not depend on
$y^*$ because the
equations ${\delta S/\delta z^A}\mid_{z^*=0}\equiv {\delta T/\delta
z^A=0}$ do not
involve the antifields $y^*_i$ or their derivatives. Generalized
auxiliary fields occur in the
transition from the total action to the extended action of the
Hamiltonian formalism
\cite{Dresse,Henneaux2}. They have properties quite similar to
ordinary auxiliary fields. In particular,
the relations that replace (\ref{11.4}) and (\ref{11.5}) are respectively
\begin{eqnarray}
S[y,y^*,z,z^*=0]=\bar S [y,y^*]+T^\prime[y,y^*,z]\\
{\delta T\over\delta y^i(x)}=\int \mu^A_i(x,x^\prime){\delta T\over
z^A(x^\prime)}dx^\prime\nonumber\\
{\delta T\over\delta y^*_i(x)}=\int \nu^{Ai}(x,x^\prime){\delta
T\over z^A(x^\prime)}dx^\prime
\end{eqnarray}
where $\bar S[y,y^*]$ is the solution of the master equation for the
reduced theory obtained by
setting $z^*_A=0$ and eliminating $z^A$ through (\ref{11.14}).

Because the equations (\ref{11.14}) are algebraic in $z^A$, one finds
again that $H_k(\delta^\prime|d)=0$
for $k>0$, where $\delta^\prime$ is now defined through
\begin{eqnarray}
\delta^\prime z^*_A={\delta S\over \delta z^A}\mid_{z^*=0},\qquad
\delta^\prime (everything\ else)=0.
\end{eqnarray}
The standard methods of homological perturbation then enable one
to establish

\begin{theorem} {\em{\bf :}}\label{11.2}
the generalized auxiliary fields do not modify $H_k(s|d)$. Namely,
$H_k(\bar s|d)\simeq H_k(s|d)$,
where (i) $s$ is the BRST differential for the formulation with $z$
and $z^*$ present~;
and (ii) $\bar s$ is the BRST differential for the formulation in which
the fields $z$ and $z^*$ are eliminated through ${\delta S/\delta
z^A}\mid_{z^*=0}=0$, $z^*_A=0$.
\end{theorem}

Theorems \ref{11.1} and \ref{11.2} imply in particular that $H(s|d)$
is invariant under transition
to the Hamiltonian formalism, provided that the inverse
transformation that expresses the
momenta and the Lagrange multipliers in terms of the velocities is local
\cite{Henneaux2}
(in space - it is of course always local in time).

Both theorems are valid in the space of infinite formal series in the
antighost number.
For the auxiliary fields that usually occur in practise, one may
improve the results as follows.
Standard auxiliary fields appear quadratically in the action, with
coefficients that depend only on the fields but not on their
derivatives, and which may be assumed to be constant under redefinition,
\begin{eqnarray}
T={1\over 2} (z^A-Z^A)(z^B-Z^B)g_{AB}.
\end{eqnarray}
The redefinition $z^A\rightarrow z^{\prime A}=z^A-Z^A(y,\partial
y,\dots)$ - which may be completed
to a canonical transformation - enables one to write $T$ as
\begin{eqnarray}
T={1\over 2} z^{\prime A}z^{\prime B}g_{AB}.
\end{eqnarray}
It is then straightforward to verify
that the solutions of the master
equations of the reduced and the unreduced theories are related as
\begin{eqnarray}
S=\bar S +T
\end{eqnarray}
and that the BRST invariant function(al)s may be taken to coincide~;
\begin{eqnarray}
A=\bar A.
\end{eqnarray}
Hence if $\bar S$ (respectively $\bar A$) is polynomial in the
antighost number, then so is $S$
(respectively $A$) and vice versa.

\section{Conclusion}

In this paper, we have derived some general theorems on the local
BRST cohomological
groups $H^k(s|d)$. We have established their link with the groups
$H_k(\delta|d)$, which are in turn connected to the groups
$H_0^k(d|\delta)$ of
$k$-forms that are closed when the equations of motion hold.
These groups are of interest in the study of the dynamics of the
theory
and have already been discussed from that point of view in the
mathematical
literature (``characteristic cohomology"). Our work makes thus a
bridge between the
local BRST cohomology and the characteristic cohomology.

We have also developed tools for calculating explicitely the groups
$H_k(\delta|d)$ for $k>1$. These tools include a vanishing theorem
for $H_k(\delta|d)$ whenever $k$ is strictly greater than the Cauchy order
of the theory. By a perturbative argument, we have then proved that
$H_2^n(\delta|d)$ vanishes for Yang-Mills theory and Einstein gravity.
This theorem is equivalent to the absence of a non trivial $2$-form
that is closed modulo the equations of motion.

In a companion paper, we shall illustrate the usefulness of the
theorems demonstrated here by computing explicitely all the
cohomological groups $H^k(s|d)$ in Yang-Mills theory, with sources included.

\section{Acknowledgements}

We are grateful to J. Collins, M. Dubois-Violette, N. Kamran, A. Slavnov, J.
Stasheff,
R. Stora, M. Talon, C. Teitelboim and C. Viallet for fruitful discussions.
This work has been supported in part by research funds from the
Belgian ``F.N.R.S."
as well as by research contracts with the Commission of the European
Communities. M.H. is grateful to the CERN theory division for
its kind hospitality while this work was being
completed.

\vfill
\eject

\end{document}